\documentclass[secnumarabic,12pt,tightenlines,eqsecnum,floats,aps,amsmath,amssymb,nofootinbib,superscriptaddress,showpacs,prd]{revtex4-1}

\usepackage{amsmath,amsthm,amssymb,amsfonts}
\usepackage{graphicx}
\usepackage[mathcal]{euler}

\numberwithin{equation}{section}

\begin{document}

\title{Quantization of Midisuperspace Models}

\author{J. Fernando \surname{Barbero G.}}
\affiliation{Instituto de
Estructura de la Materia, CSIC, Serrano 123, 28006 Madrid, Spain}

\author{Eduardo J. \surname{S. Villase\~nor}}
\affiliation{Instituto Gregorio
Mill\'an, Grupo de Modelizaci\'on y Simulaci\'on Num\'erica,
Universidad Carlos III de Madrid, Avda. de la Universidad 30, 28911
Legan\'es, Spain} \affiliation{Instituto de Estructura de la
Materia, CSIC, Serrano 123, 28006 Madrid, Spain}

\date{October 1, 2010}

\begin{abstract}
We give a comprehensive review of the quantization of midisuperspace
models. Though the main focus of the paper is on  quantum aspects, we
also provide an introduction to several classical points related to
the definition of these models. We cover some important issues, in
particular, the use of the principle of symmetric criticality as a
very useful tool to obtain the required Hamiltonian formulations. Two
main types of reductions are discussed: those involving metrics with
two Killing vector fields and spherically symmetric models. We also
review the more general models obtained by coupling matter fields to
these systems. Throughout the paper we give separate discussions for
standard quantizations using geometrodynamical variables and those
relying on loop quantum gravity inspired methods.
\end{abstract}

\maketitle

%================================================================
\newpage

\tableofcontents

\newpage

\section{Introduction}
\label{sect1}

The problem of quantizing general relativity (GR) is a very hard
one. To this day, and despite continuing efforts, there is not a
completely satisfactory theory of quantum gravity. In order to acquire
the necessary intuition to deal with the very many issues whose resolution
is required, it is often useful to concentrate on simplified models that
exhibit only \textit{some} of the difficulties present in the full theory
(and hopefully in a milder guise).

A natural way to get simplified versions of a physical system is to
introduce symmetries. They often allow us to get particular solutions
in situations where a complete resolution is out of reach. Usually
this happens because the effective dimensionality of the problem is
reduced by the symmetry requirements. In the case of classical
(i.e.\ non-quantum) systems, symmetry is often used at the level of the
equations of motion. Actually, a good way to start exploring the
concrete physics of a given model is to look for symmetric
configurations. The effectiveness of this approach is reinforced by
the fact that, in many instances, they are good approximations to real
physical situations. For example, the waves produced on a water pond
by a falling stone are very well described by rotationally invariant
functions satisfying the two-dimensional wave equation. This is so
because the initial disturbance producing the wave is rotationally
symmetric to a good approximation. This very same philosophy is used
in many branches of Physics. In GR, for instance, the most important
and useful metrics solving the Einstein field equations exhibit some
type of symmetry -- just think of the Schwarzschild, Kerr, or Friedman
spacetimes. Actually only a few closed form solutions to the Einstein
field equations with no Killing fields are
known~\cite{Stephani:2003tm}.

There is a vast literature devoted to the classical aspects of the
symmetry reductions that covers topics ranging from purely
mathematical issues to physical applications in the fields of
cosmology and black hole physics. These simplified systems also
provide interesting quantum theories that are easier to handle than
full gravity. There are two main types of models that can loosely be
defined as those with a finite number of degrees of freedom
(minisuperspaces) and those that require the introduction of
infinitely many of them (midisuperspaces). The purpose of this Living
Review is to explore the quantization of the latter, hence, we will
only discuss those classical aspects that are of direct relevance for
their quantization (for example the Hamiltonian description).

\bigskip

The paper is organized as follows. After this introduction we will
review the history of midisuperspaces in Section~\ref{sect2}. To this
end we will give a general overview of symmetry reductions of GR. A
very important idea that plays a central role in this subject is the
\textit{principle of symmetric criticality}. It provides a very useful
simplification -- especially when considering the Hamiltonian
framework -- because it allows us to derive everything from a
symmetry-reduced variational principle obtained by restricting the
Einstein--Hilbert action to the family of symmetric solutions of
interest. Though not every reduced system is of this type this happens
to be the case for all the models that we will consider in the
paper. After a discussion on some aspects concerning the mathematical
description of superspace we will comment on the differences between
minisuperspaces and midisuperspaces.

General issues concerning quantization will be addressed in
Section~\ref{sect3}, where we will quickly review -- with the idea of
dealing with the quantization of symmetry reductions -- the different
approaches to the quantization of constrained systems, i.e.\ reduced
phase space quantization, Dirac quantization, quantization of fully
and partially gauge fixed models and path integral methods. We will
end this section with a discussion of the differences between the
``quantizing first and then reducing'' and the ``reducing first and
then quantizing'' points of view.

Section~\ref{sect4} is devoted to the discussion of some relevant
classical aspects of midisuperspaces. We will consider, in particular,
one-Killing vector reductions, two-Killing vector reductions and
spherically symmetric models and leave to Section~\ref{sect5} the main
subject of the paper: the quantization of midisuperspaces. There we
will review first the one-Killing vector case and then go to the more
important -- and developed -- two-Killing vector reductions for which we
will separately consider the quantization of Einstein--Rosen waves,
Gowdy cosmologies and other related models. A similar discussion will
be presented for spherically symmetric midisuperspaces. We will look
at both metric and loop quantum gravity (LQG) inspired quantizations
for the different models. We conclude the paper with our conclusions
and a discussion of the open problems.

\bigskip

We want to say some words about the philosophy of the paper. As the
reader will see there is an unusually low number of formulas. This is
so because we have chosen to highlight the main ideas and emphasize
the connection between the different approaches and models. We provide
a sizable bibliography at the end of the paper; technical details can
be found there. We feel that the proper way to master the subject is
to read the original papers, so we believe that we will have reached
our goal if the present work becomes a useful guide to understand the
literature on the subject. We have tried to give proper credit to all
the researchers who have made significant contributions to the
quantization of midisuperspaces but of course some omissions are
unavoidable. We will gladly correct them in coming updates of this
Living Review.

%================================================================
\newpage

\section{A Minihistory of Midisuperspaces}
\label{sect2}

\subsection{Symmetry reductions in classical and quantum general relativity}

Classical systems are (usually) described in terms of field models
whose dynamics is given by partial differential equations derived from
a variational principle. A symmetry reduced model associated with a
given classical system is defined as one obtained by considering only
those solutions to the equations of motion that satisfy a certain
symmetry condition. In order to describe these reduced models one has
to follow several steps (see~\cite{Torre:1998dy} for a careful and
complete discussion of these issues):

\begin{itemize}
 \item Defining a group action on the solution space of the full
 model.
 \item Finding a suitable parametrization of the solutions invariant
 under the group action.
 \item Obtaining the field equations describing the symmetric
 configurations.
\end{itemize}

When these steps can be successfully carried out, the final outcome of
this process is a set of equations for the symmetry reduced
system. There are two conceivable ways to get them. The direct one
consists in particularizing the general field equations to the
invariant solutions obtained in the second step (by using some of the
parametrizations introduced there). A second more indirect way would
rely on the use a symmetry reduced action principle. This may seem as
an unnecessary detour but, if we intend to quantize the reduced system,
it becomes an unavoidable step as we need a Hamiltonian formulation to
define the dynamics of the quantized model. Though one may naively
expect that the reduced action can be obtained by just restricting the
one describing the full (i.e.\ non-reduced) model to the parameterized
symmetric configurations, there are subtleties that may actually
prevent us from doing so. We will discuss these problems in
Section~\ref{ssect211} devoted to symmetric criticality.

The transit to the quantum version of symmetry reductions of classical
theories (involving either mechanical systems or fields) is quite
non-trivial. This is a very important topic that plays a central role
in the present paper so we discuss it here in some detail. There are
several questions to be addressed in this respect:

\begin{itemize}
\item Definition of the symmetric quantum states and/or quantum
  symmetry reductions.
\item Evolution of the symmetric states under the full dynamics and
  the reduced dynamics.
\item Comparison between the two: can we derive one from the other?
\end{itemize}

The first of these issues is usually discussed as the problem of
understanding the commutativity of symmetry reduction and
quantization, i.e.\ to figure out if the result of ``first quantizing
and then reducing'' is the same as the one of ``first reducing and
then quantizing''. The other two items are also important, for
example, to assess to what extent the results obtained in quantum
cosmology (in its different incarnations including loop quantum
cosmology) can be taken as hard physical predictions of quantum
gravity and not only as suggestive hints about the physics of the
complete theory. Of course the usual problems encountered in the
quantization of constrained systems will be also present here. We will
return to these issues in Section~\ref{sect3}.

\subsubsection{Symmetric criticality}
\label{ssect211}

The original formulation of the principle of symmetric criticality,
telling us when symmetric extremals of a functional can be obtained as
the ones corresponding to the symmetry reduction of it, was stated by
Palais in a variety of different settings~\cite{Palais:1979}. The
adaptation of this principle to general relativity was discussed in
detail by Fels and Torre~\cite{Fels:2001rv} though its importance was
recognized since the early seventies (see \cite{Isenberg:1980} for
an excellent review).

As mentioned above, the classical reduction process for a field theory
is performed in several steps~\cite{Palais:1979,Torre:1998dy}. One starts
by defining a group action on the space of fields of the model, find
then a parametrization of the most general configuration invariant
under the group action and, finally, obtaining the form of the
equations of motion restricted to these symmetric field configurations
(the \textit{reduced field equations}). General solutions to these
equations correspond to symmetric solutions of the full theory.

In the case of general relativity one can ask oneself if the reduced
field equations can be obtained as the Euler--Lagrange equations
derived from some \textit{reduced Lagrangian} and also if this
Lagrangian can be obtained by simply restricting the Einstein--Hilbert
action to the class of metrics compatible with the chosen
symmetries. Obviously this would be the simplest (and more desirable)
situation but one cannot exclude, in principle, that the reduced field
equations could come from an action that is not the symmetry reduced
one (or even that they cannot be derived from a well defined action
principle). If a Hamiltonian formulation can be obtained for a
symmetry reduction of a physical system then it is possible to
consider its quantization. This is the path followed in quantum
cosmology and in the study of the midisuperspace models that are the
subject of this review.

The parametrization of the invariant field configurations usually
involves the introduction of a set of arbitrary functions whose number
is smaller than the number of original field components. Furthermore,
a judicious choice of coordinates adapted to the symmetry, normally
restricts the number of \textit{variables} upon which these functions
depend. In some instances it is possible to work with a single
independent variable. This happens, for example, in Bianchi models
where these unknown functions depend only on a ``time coordinate" that
labels compact homogenous spatial slices of spacetime. Another
instance of this behavior is provided by \textit{static}, spherical,
vacuum space-times where the arbitrary functions appearing in the
metric depend on an area variable (usually proportional to $r^2$). In
both cases the field equations reduce to ordinary differential
equations. This, in turn, shows that these particular symmetry
reductions of general relativity describe systems with a finite number
of degrees of freedom; i.e.\ purely mechanical models.

Necessary and sufficient conditions guaranteeing that the principle of
symmetric criticality holds in general relativity are given in
Theorem~5.2 of reference~\cite{Fels:2001rv}. They are technical in
nature but their role is to prevent the occurrence of the two
conceivable scenarios in which the symmetric criticality principle may
fail. The first has to do with the possibility that the surface terms
coming from integration by parts after performing variations in the
full action do not reduce to the ones corresponding to the reduced
action (this is what happens for Bianchi B models). The second is
related to the fact that considering only ``symmetric variations'' may
not give all the field equations but only a subset of them. An
important comment to make at this point is that it is always possible
to check if the symmetric criticality principle holds just by considering
the group action because it is not tied to the form of a specific
Lagrangian. This remarkable fact allows us to check the validity of the
principle for whole families of symmetric models irrespective of their
dynamics. In fact, for the types of vacuum models that are the main
subject of this Living Review, symmetric criticality can be shown to
hold~\cite{Fels:2001rv, Torre:1998dy} and, hence, we have a simple way
to get a Hamiltonian for the reduced systems. In the spherically symmetric
case the result holds as a consequence of the compactedness of the group
of symmetries~\cite{Torre:1998dy} (in the case of the two-Killing
symmetry reductions the validity of the principle is justified in the
papers~\cite{Fels:2001rv, Torre:1998dy}). If scalar fields are coupled
to gravity the principle still holds, however the introduction of other
matter fields must be treated with care because their presence may
influence the action of the symmetry group \cite{Fels:2001rv}.

\subsection{Superspace}

Wheeler's notion of superspace is inextricably linked to the problem
of understanding quantum general relativity. In a nutshell superspace
can be defined as the space of geometries for the three dimensional
manifolds that constitute \textit{space} in the dynamical picture of
GR that we have come to know as geometrodynamics. As the study of
symmetry reductions require us to restrict the possible configurations
to a subset of the full superspace it is convenient to discuss, at
least briefly, some of its basic features.

Superspace plays the role of the configuration space for general
relativity in the traditional metric representation. The associated
cotangent bundle, when properly defined, is the phase space for the
Hamiltonian formulation of the theory. As a Hamiltonian formulation is
the starting point for the quantization of any mechanical or field
system, the role of superspace and the need to understand its
mathematical structure cannot be overemphasized. A secondary role of
superspace is that of providing ``variables for the wave function'' in
a functional Schr\"{o}dinger representation for quantum
gravity. However, it should be noted at this point that even in the
quantization of the simplest field theories -- such as scalar fields --
it is necessary to suitably enlarge this configuration space and allow
for distributional, non-smooth, objects to arrive at a consistent
model (see, for example,~\cite{Ashtekar:2004eh}). How -- and if -- this
can be done in the geometrodynamical setting is an interesting, if
hard, question. This is directly related to the Wheeler--DeWitt
approach to the quantization of GR~\cite{DeWitt:1967yk}.

The precise definition of the geometry of a three manifold requires
some discussion (see~\cite{Fischer, Giulini:2009np} and references
therein for a nice introduction to the subject). Here we will content
ourselves with a quick review of the most important issues. It is
important to remark at this point that the non-generic character of
geometries with non-trivial isometry groups has a very clear
reflection in superspace: they correspond to singularities.

The \textit{geometry} of a four dimensional manifold in the
relativists parlance refers to equivalence classes of suitably smooth
Lorentzian metrics defined on it. Two metrics are declared equivalent
if they are connected by a diffeomorphism. Though one might naively
think that this is just a mathematically sensible requirement, in
fact, it is quite natural from a physical point of view. The reason is
that ultimately the geometry must be probed by physical means. This,
in turn, demands an operational definition of the (possibly idealized)
physical processes allowing us to explore -- actually measure --
it. This is in the spirit of special and general relativity where the
definition of physical magnitudes such as lengths, distances,
velocities and the like requires the introduction of concrete
procedures to measure them by using basic tools such as clocks, rulers
and light rays. Every transformation of the manifold (and the objects
defined on it) that does not affect the operational definition of the
measuring processes will be physically unobservable. Diffeomorphisms
are such transformations. Notice that this prevents us from
identifying physical events with points in the spacetime manifold as a
diffeomorphism can take a given event from one point of the manifold
to another (see~\cite{Misner:1957wq}).

The precise definition and description of the space of geometries
requires the introduction of mathematical objects and structures at
different levels:

\begin{itemize}
\item The three-dimensional \textit{smooth} manifold $\Sigma$ with the
  required differential structure.
\item A class of smooth Riemannian metrics on this manifold that
  should be considered (smooth in the sense of being
  $C^\infty(\Sigma)$). We will denote this as $\mathrm{Riem}(\Sigma)$.
\item A suitable topology and differential structure on this space.
\item An equivalence relation between different smooth metrics
  provided by the action of (a class of) \textit{smooth} ($C^\infty$)
  diffeomorphisms $\mathrm{Diff}(\Sigma)$ on $\mathrm{Riem}(\Sigma)$.
\end{itemize}

After doing this one has to study the quotient
$\mathrm{Riem}(\Sigma)/\mathrm{Diff}(\Sigma)$. Naturally, superspace
will inherit some \textit{background} properties from those carried by
the different elements needed to properly define it. The resulting
space has a rich structure and interesting properties that we will
very quickly comment here (the interested reader is referred
to~\cite{Giulini:2009np} and the extensive bibliography cited
there).

An important issue is related to the appearance of singularities in
this quotient space associated with the fact that in many instances the
spatial manifold $\Sigma$ allows for the existence of invariant
metrics under non-trivial symmetry groups (leading to a non-free
action of the diffeomorphisms). This turns out to be a problem that
can be dealt with in the sense that the singularities are minimally
resolved (see~\cite{Fischer}). It is important to mention at this
point that the symmetry reductions that we will be considering here
consist precisely in restrictions to families of symmetric metrics
that, consequently, sit at the singularities of the full
superspace. This fact, however, does not necessarily imply that the
reduced systems are pathological. In fact some of them are quite well
behaved as we will show in Section~\ref{sect4}.

Finally we point out that both the space of Riemannian metrics
$\mathrm{Riem}(\Sigma)$ and the quotient space mentioned above are
endowed with natural topologies that actually turn them into very well
behaved topological spaces (for instance, they are metrizable -- and
hence paracompact --, second countable and connected). The space of
metrics $\mathrm{Riem}(\Sigma)$ can be described as a principal bundle
with basis $\mathrm{Riem}(\Sigma)/\mathrm{Diff}_\infty(\Sigma)$ and
structure group given by $\mathrm{Diff}_\infty(\Sigma)$ (that is, the
proper subgroup of those diffeomorphisms of $\Sigma$ that fix a
preferred point $\infty\in \Sigma$ and the tangent space at this
point). Finally a family of ultralocal metrics is naturally defined in
superspace~\cite{Giulini:2009np}. Some of these properties are
inherited by the spaces of symmetric geometries that we consider here.

Other approaches to the quantization of GR, and in particular loop
quantum gravity, rely on spaces of connections rather than in spaces
of metrics. Hence, in order to study symmetry reductions in these
frameworks, one should discuss the properties of such ``connection
superspaces'' and then consider the definition of symmetric
connections and how they fit into these spaces. The technical
treatment of the spaces of Yang--Mills connections modulo gauge
transformations has been developed in the late seventies by Singer and
other authors~\cite{Singer,Mitter}. These results have been used by
Ashtekar, Lewandowski~\cite{Ashtekar:1993wf} and others to give a
description of the spaces of connections modulo gauge (encompassing
diffeomorphisms) and their extension to symmetry reductions have been
explored by Bojowald~\cite{Bojowald:1999eh} and collaborators as a
first step towards the study of symmetry reductions in LQG. These will
be mentioned in the last section of the paper.

\subsection{Minisuperspaces}

Minisuperspaces appear when the symmetry requirements imposed upon
spacetime metrics are such that the dimension of
$\mathrm{Riem}(\Sigma)$ (and, hence, of
$\mathrm{Riem}(\Sigma)/\mathrm{Diff}(\Sigma)$) becomes finite.
Historically these were the first symmetry reductions of general
relativity that received serious consideration from the quantum point of
view~\cite{DeWitt:1967yk,DeWitt:1967ub,DeWitt:1967uc,Misner:1972js}.
Their main advantage in the early stages of the study of quantum
gravity was the fact that the resulting models were finite-dimensional
and their quantization could be considered in a more or less
straightforward way. Important conceptual problems received attention
within this setting; in particular those related to the interpretation
of the universe wave function and the resolution of cosmological
singularities. They are receiving renewed attention these days as very
useful test beds for loop quantum gravity (the so called loop quantum
cosmology or LQC in short). This is so both at the technical level and
regarding physical predictions. In particular the resolution of the
initial singularity in LQC is a tantalizing hint of the kind of
fundamental knowledge about the universe that a complete theory of
quantum gravity could provide.

The Bianchi models are arguably the most important among the
minisuperspaces. They describe spatially homogeneous (but generally
non-isotropic) cosmologies. These spacetimes are obtained
(see~\cite{Ryan:1975jw, Wald:1984rg} for a pedagogical presentation) by
requiring that the space-time admits a foliation by smooth
three-dimensional hypersurfaces $\Sigma_t$ that are the orbits of a
group of isometries $G$. When the action $G$ is required to be simply
transitive (i.e.\ for each pair of points $p,q\in\Sigma_t$ there exist
a unique element of $g\in G$ such that $g\cdot p=q$) its dimension
must be 3. In addition to the Bianchi models there are other spatially
homogeneous spacetimes for which the group action is not simply
transitive (or does not have a subgroup with a simply transitive action).
These are the so called Kantowski-Sachs models with
$G=\mathbf{R}\times SO(3)$ and such that the spatial homogeneous
hypersurfaces are $\mathbf{R}\times S^2$. Metrics for the Bianchi models
are parameterized by functions of the ``time'' variable that labels the
sheets of the spacetime foliation and can be conveniently written by
using a basis of invariant one forms. The Killing vector fields of the
metric induced on each $\Sigma_t$ are in one to one correspondence with
the right invariant vector fields in the group $G$ and satisfy the
commutation rules of the Lie algebra of $G$.

The Einstein field equations reduce in these cases to a system of
ordinary differential equations. Bianchi models are classified as type
A and type B depending on some invariant properties encoded in the
structure constants $C_{ab}^{\,\,\,\,c}$ of the isometry group. If
they satisfy the condition $C_{ab}^{\,\,\,\,b}=0$ the resulting model
is type A, otherwise it is called type B. Only the type A ones satisfy
the principle of symmetric criticality and can be quantized in a
straightforward way~\cite{Fels:2001rv}.

Two main approaches are possible to study the classical dynamics of
minisuperspace models and, in particular the Bianchi models: The
covariant spacetime textbook approach (see, for
example,~\cite{Wald:1984rg}) that directly looks for the spatially
homogeneous solutions to the Einstein field equations, and the
Hamiltonian one that can be applied when the principle of symmetric
criticality holds. Of course they are ultimately equivalent but the
descriptions that they provide for the classical dynamics of these
systems are surprisingly different. A very good account of these
issues can be found in~\cite{Ashtekar:1991wa}. Among the points that
are worthwhile singling out maybe the most striking one refers the
identification and counting of the number of degrees of freedom. As it
can be seen, these numbers generically disagree in the case of open
spatial slices. This can be easily shown~\cite{Ashtekar:1991wa} for
the Bianchi~I model for $\mathbf{R}^3$ spatial slices. From the
covariant point of view the family of solutions of Bianchi type I is
fully described by a single parameter; on the other hand the
Hamiltonian analysis (this is a constrained Hamiltonian system) tells
us that the number of phase space degrees of freedom is ten
corresponding to five physical degrees of freedom. The resolution of
this problem~\cite{Ashtekar:1991wa} requires a careful understanding
of several issues:

\begin{itemize}

\item The role of the spatial topology. It can be seen that this
  mismatch does not occur for compact spatial topologies (which, by
  the way, are impossible for the Bianchi type~B case). In this case
  the appearance of global degrees of freedom reconciles the covariant
  and Hamiltonian points of view.

\item The difference between gauge symmetries and non-gauge
  symmetries. The first ones are those generated by the constraints in
  the Hamiltonian formulation -- such as the familiar $U(1)$ gauge
  invariance of electromagnetism -- and connect physically
  indistinguishable configurations. The non-gauge ones correspond to
  homogeneity preserving diffeomorphisms that connect physically
  distinct solutions. They are not generated by constraints. A
  standard example is the Poincar\'e invariance in standard quantum
  field theories.

\item The need to understand the different roles played by
  diffeomorphisms in the spacetime and Hamiltonian pictures. Whereas
  from the spacetime point of view solutions to the Einstein equations
  that can be connected by the action of a diffeomorphism are
  considered to be physically equivalent they may not be so from the
  Hamiltonian point of view in which a spacetime foliation must be
  introduced.

\end{itemize}

The bottom line can be summarized by saying that the extra structure
present in the Hamiltonian framework provides us with sharper tools to
separate gauge and symmetries than the purely geometric point of view
of the standard covariant approach~\cite{Ashtekar:1991wa}. If one is
interested in the quantization of these minisuperspace reductions the
Hamiltonian framework is the natural (and essentially unavoidable)
starting point.

It is obvious that essentially all the points discussed here will be
relevant also in the case of midisuperspaces, though to our knowledge
the current analyses of this issue are far from complete -- and
definitely much harder -- because one must deal with infinite dimensional
spaces. In this case, as we will see, the gauge symmetry remaining
after the symmetry reduction will include a non-trivial class of
restricted diffeomorphisms. This is, in fact, one of the main reasons
to study these symmetry reductions as they may shed some light on the
difficult issue of dealing with diffeomorphism invariance in full
quantum gravity. A final interesting point that we want to mention is
the problem of understanding how minisuperspace models sit inside the
full superspace. This has been discussed by Jantzen
in~\cite{Jantzen:1979}.

\subsection{Midisuperspaces}

The type of phenomena that can be described by a minisuperspace is
rather limited because the metrics in these models effectively depend
on a finite number of parameters. A less drastic simplification would
consist in allowing some functional freedom but not the most general
one. This is in essence the definition of a midisuperspace. More
specifically the idea is to impose again symmetry requirements to
restrict the set $\mathrm{Riem}(\Sigma)$ used in the superspace
construction in such a way that the allowed metrics are parametrized
by \textit{functions} rather than by numerical parameters. By doing this
the hope is to increase the number of degrees of freedom of the models
and eventually have local degrees of freedom. Notice that, as we will
discuss below, the presence of fields at this stage does not preclude the
possibility of having a finite dimensional reduced phase space.

This can be accomplished, in particular, by restricting ourselves to
metrics having a ``low'' number of spatial Killing vector fields. As
we will see in the following, the case in which spacetime metrics are
required to have two commuting Killing vector fields is specially
appealing because some of these models are solvable both at the
classical and quantum levels while, on the other hand, it is possible
to keep several interesting features of full general relativity such
as an infinite number of degrees of freedom and diffeomorphism
invariance. The Einstein--Rosen waves (ER)~\cite{Einstein:1937qu,Beck}
were the first symmetry reduction of this type that was considered
from the Hamiltonian point of view with the purpose of studying its
quantization~\cite{Kuchar:1971xm}. As a matter of fact, Kucha\v{r}
introduced the term \textit{midisuperspace} precisely to refer to this
system~\cite{Kuchar:1971xm,Kuchar:1973}. Other configurations of this
type are the well-known Gowdy spacetimes~\cite{Gowdy:1971jh,Gowdy:1973mu}
that have been used as toy models in quantum gravity due to their
possible cosmological interpretation.

A different type of systems that have been extensively studied and
deserve close investigation are the spherically symmetric ones (in
vacuum or coupled to matter). These are, in a sense, midway between the
Bianchi models and the midisuperspaces with an infinite number of
\textit{physical} degrees of freedom such as ER. General spherically
symmetric spacetime metrics depend on functions of a radial coordinate
and time so these models \textit{are field theories}. On the other
hand, in vacuum, the space of physically different spherical solutions
to the Einstein field equations is finite dimensional (as shown by
Birkhoff's theorem). This means that the process of finding the
reduced phase space (or, alternatively, gauge fixing) is
non-trivial. The situation usually changes when matter is coupled
owing to the presence of an infinite number of matter degrees of
freedom in the matter sector. The different approaches to the
canonical quantization of these types of models is the central topic
of this Living Review.

%================================================================
\newpage

\section{Quantization(s)}
\label{sect3}

The canonical treatment of the symmetry reductions of general
relativity requires the understanding of constrained Hamiltonian
systems. In the cases that we are going to discuss (and leaving aside
functional analytic issues relevant for field
theories~\cite{Gotay:1978}), the starting point consists of the
classical unconstrained configuration space $\mathcal{C}$ of the model
and the cotangent bundle $\Gamma$ over $\mathcal{C}$ endowed with a
suitable symplectic form $\Omega$. A dynamical Hamiltonian system is
said to be constrained if the physical states are restricted to belong
to a submanifold $\bar{\Gamma}$ of the phase space $\Gamma$, and the
dynamics is such that time evolution takes place within
$\bar{\Gamma}$~\cite{Gotay:1978}. In the examples relevant for us the
space $\bar{\Gamma}$ will be globally defined by the vanishing of
certain sufficiently regular constraint functions, $C_I=0$. In the case
of general relativity these constraint functions are the integrated
version of the scalar and vector constraints and the subindex $I$ refers
to lapse and shift choices (see, for example,~\cite{Ashtekar:1982wv}).
Notice, however, that there exist infinitely-many constraint equations
that define \textit{the same} submanifold $\bar{\Gamma}$. The choice of
one representation or another is, in practice, dictated by the
variables used to describe the physical system. We will assume that
$\bar{\Gamma}$ is a first class submanifold of $\Gamma$. This is
geometrical property that can be expressed in terms of the concrete
constraint equations describing $\bar{\Gamma}$ as
\begin{equation}
\left.\{C_I,C_J\}\right|_{\bar{\Gamma}}=0\,.
\end{equation}
The pull-back of the symplectic structure of $\Gamma$ to
$\bar{\Gamma}$ is degenerate and the integral submanifolds defined by
the degenerate directions are the so called \textit{gauge orbits}. The
reduced phase space $\tilde{\Gamma}$ is the quotient space whose
points are the orbits of the gauge flows. It can be endowed with the
natural symplectic structure $\tilde{\Omega}$ inherited from
$\Omega$. If a non trivial dynamics describes the evolution of the
system in $\tilde{\Gamma}$ this will be given by the reduced
Hamiltonian $\tilde{H}$ obtained by restricting the original one to
$\tilde{\Gamma}$. This restriction is well defined whenever $H$ is
gauge invariant and, hence, constant on the gauge orbits.

\subsection{Reduced phase space quantization}

The reduced phase space quantization is simply the quantization of the
reduced space $(\tilde{\Gamma},\tilde{\Omega}, \tilde{H})$ of the
constrained Hamiltonian system whenever this is possible. Notice that
the process of taking quotients is highly non-trivial and many
desirable regularity properties need not be preserved. In the models
that we consider in this paper we will suppose that no obstructions
appear so that the reduced phase space is well defined. Even in this
case some difficulties may (and in practice do) arise, in particular:

\begin{itemize}

\item The characterization of the quotient space $\tilde{\Gamma}$ is
  usually very difficult even when the quotient itself is well
  defined. In practice this reduced phase space is effectively
  described by using a gauge fixing procedure that picks a single
  field configuration from each gauge orbit in a smooth way (whenever
  this is possible).

\item In general, there is no guarantee that $\tilde{\Gamma}$ will be
  the cotangent space of a reduced configuration manifold
  $\tilde{\mathcal{C}}$. Although there are techniques that may allow
  us to tackle this situation (i.e.\ geometric
  quantization~\cite{Woodhouse:1992pa}) they are not always
  straightforward to apply.

\item It may be difficult to extract physics from the reduced phase
  space description. In practice, even when the reduction can be
  carried out in an explicit way, it is very difficult to reexpress
  the results in terms of the original variables in which the problem
  is naturally written.

\end{itemize}

The reduced phase space description amounts to the identification of
the true physical degrees of freedom of the system. As a rule, for
many physical models (an certainly in the case of gravitational
theories) this description is either unavailable or extremely
difficult to handle. In these cases one is forced to learn how to live
with the redundant descriptions provided by gauge theories and how to
handle the constraints both at the classical and quantum
levels. Finally it is important to notice that whenever the reduced
phase space can be characterized by means of a gauge fixing the
quantization ambiguities that may arise do not originate in the
different gauge choices -- as long as they are acceptable -- but
rather in the possibility of having different quantizations for a
given classical model. This is so because from the classical point of
view they are explicit representations of the same abstract object:
the reduced phase space~\cite{Henneaux:1992ig}. There are several
approaches to the quantization of gauge systems that we will briefly
discuss next.

\subsection{Dirac quantization}

In Dirac's approach to quantization one starts from a kinematical
vector space $V$ adapted (i.e.\ with the right dimensionality among
other requirements) to the description of a physical system defined on
the phase space $\Gamma$. The constraints $C_I=0$ are then represented
as operators whose kernels define the physical states $\Psi\in V$ of
the quantum theory, $\hat{C}_I\Psi=0$. Finally, to define
probabilities, the physical subspace $V_{\mathrm{phys}}$ is endowed with a
suitable inner product $\langle\cdot,\cdot\rangle$ such that
$(V_{\mathrm{phys}},\langle\cdot,\cdot\rangle)$ becomes a Hilbert space
$\mathcal{H}_{\mathrm{phys}}$. In order to make these ideas explicit, the
following concrete points must be addressed:

\begin{itemize}

\item The identification of a set of elementary phase space variables
  for the full (non-constrained) phase space of the classical system.

\item The selection of a suitable Poisson algebra on the full phase
  space generated by the elementary variables.

\item The construction of a representation for this Poisson
  algebra on the complex vector space $V$.

\item The implementation of the first class constraints $C_I= 0$ as
  operators acting on the representation space.

\item The characterization of the physical states, i.e.\ the space
  $V_{\mathrm{phys}}$ spanned by those vectors in the kernel of all the
  constraint operators.

\item The identification of physical observables (the operators that
  leave $V_{\mathrm{phys}}$ invariant).

\item Finally, if we want to answer physical questions -- such as
  probability amplitudes or expectation values -- we need to endow
  $V_{\mathrm{phys}}$ with an Hermitian inner product.

\end{itemize}

The Wheeler--DeWitt approach and loop quantum gravity both follow the
spirit of the Dirac quantization of constrained systems mentioned
here. In LQG~\cite{Ashtekar:2004eh}, the kinematical vector space $V$
is endowed with a Hilbert space structure defined in terms of the
Ashtekar-Lewandowski measure. However, the identification of the inner
product in the space of physical states is not as simple as the
restriction of the kinematical Hilbert structure to the physical
subspace because the spectrum of the constraint operators may have a
complicated structure. In particular it may happen that the kernel of
these operators consists only of the zero vector of the kinematical
Hilbert space. The Wheeler--DeWitt approach is less developed from the
mathematical point of view but many constructions and ideas considered
during the mathematical development of LQG can be exported to that
framework. It is important to mention that under mathematical
restrictions similar to the ones imposed in LQG some crucial
uniqueness results (specifically the LOST~\cite{Lewandowski:2005jk}
and Fleischack~\cite{Fleischhack:2004jc} theorems on the uniqueness of
the vacuum state) do not hold~\cite{Alink}. Though the approach can
probably be developed with the level of mathematical rigor of LQG this
result strongly suggest that LQG methods are better suited to reach a
complete and fully consistent quantum gravity theory. In any case we
believe that it could be interesting to explore if suitable changes in
the mathematical formulation of the Wheeler--DeWitt formalism could
lead to uniqueness results of the type already available for LQG.

\subsection{Quantization with partial gauge fixing}

As mentioned above the reduced phase space is the space of gauge
orbits endowed with a symplectic structure $\tilde{\Omega}$ inherited
from the original one $\Omega$ in the full phase space. A strategy
that is useful in the context of midisuperspaces is to partially fix
the gauge. In practice this means that the dimensionality of the
constraint hypersurface (and, as a consequence, of the gauge orbits)
is reduced. This may be useful if one is interested in leaving some
residual gauge symmetry in the model on purpose (such as radial
diffeomorphisms in spherically symmetric
models~\cite{Campiglia:2007pr}) to check if one can deal with it in
some quantization scheme. In other situations the natural gauge fixing
conditions simply fail to fix the gauge completely; this happens, for
example in the compact Gowdy models~\cite{Misner:1973zz}. In such
cases the residual gauge invariance is usually treated by employing
Dirac's procedure though other approaches are, of course, possible. A
very attractive feature of the resulting formulation is that the
quantum dynamics of the model is given by a ``time'' dependent
Hamiltonian that can be studied in great detail due to its relatively
simple structure. This is possible because its meaning can be
understood by using results developed in the study of the
time-dependent harmonic oscillator (see~\cite{Vergel:2009st} and
references therein).

\subsection{Path integral quantization}

An alternative quantization, that has been successfully employed in
standard quantum field theories, consists in using a path integral to
represent relevant physical amplitudes and then develop perturbative
techniques to extract the physical information as some kind of
expansion (usually asymptotic) in terms of coupling constants. The
main idea is to represent transition amplitudes as integrals over a
set of ``interpolating configurations'' (trajectories for particle
systems or, more generally, field histories). For example, the
expression
\begin{equation}
( g_2,\phi_2,\Sigma_2|g_1,\phi_1,\Sigma_1)=\int \exp\big(iS(g,\phi)\big) \,\mu(\mathrm{d}g,\mathrm{d}\phi)
\end{equation}
would represent the probability amplitude to go from a state with
metric $g_1$ and matter fields $\phi_1$ on a 3-surface $\Sigma_1$ to a
state with metric $g_2$ and matter fields $\phi_2$ on a 3-surface
$\Sigma_2$. Here $S$ is the classical action and $\mu$ is the
``measure'' on the space of field configurations determined by the
phase space measure. The integral has to be computed over the field
configuration on the region of the space-time manifold which has
$\Sigma_1$ and $\Sigma_2$ as boundaries. One of the advantages that
are usually attributed to the path integral is that it provides a
``covariant'' approach to quantum field theory. However, it is
important to notice that only the \textit{phase space path integral}
can be shown to be formally equivalent to canonical
quantization~\cite{Henneaux:1992ig}. If the integral in the momenta
can be performed in a closed algebraic form one gets a configuration
space path integral whose integrand can be seen to be, in some cases
but not always, just the action expressed in terms of configuration
variables and their derivatives. In the case of reduced phase space
quantization the correct writing of the path integral requires the
introduction of the so called Fadeev-Popov terms that take into
account the fact that the integration measure is the pullback of the
formal Liouville measure to the hypersurface defined by the first
class constraints and gauge fixing conditions (or alternatively to the
space of gauge orbits~\cite{Henneaux:1992ig}). The path integral
method can be rigorously defined in some instances, for example in the
quantum mechanics for systems with a finite number of degrees of
freedom and some field theories, such as topological models and
lower dimensional scalar models. In other cases, though, it is just a
formal (though arguably very useful) device.

The first proposals to use path integrals in quantum gravity go back
to the fifties (see, for example, the paper by
Misner~\cite{Misner:1957wq}) and were championed by Hawking, among
many other authors, in the study of quantum cosmology, black hole
physics and related problems. Path integrals are also useful in other
approaches to quantum gravity, in particular Regge calculus, spin
foams and causal dynamical triangulations (see the Living Review by
Loll~\cite{Loll:1998aj} and references therein). Finally they
establish a fruitful relationship between quantum field theory and
statistical mechanics.

Although the majority of the work on quantum midisuperspaces uses the
canonical approach, there are nonetheless some papers that use
standard perturbative methods based on path integrals to deal with
some of these models, for example the Einstein--Rosen
waves~\cite{Niedermaier:2002eq,BarberoG.:2003pz}. The results obtained
with these methods suggest that this model, in particular, is
renormalizable in a generalized sense and compatible with the
asymptotic safety scenario~\cite{Niedermaier:2006wt}.

\subsection{Symmetry reductions and quantization}
\label{ssect35}

Many problems in quantum mechanics reduce to the computation of
transition probabilities. For instance, in the case of a free particle
moving in three dimensions all the relevant information about the
quantum evolution can be encoded in the propagator $(x_1,t_1|x_2,t_2)$
giving the probability amplitude to find the particle at $x_2$ in the
time instant $t_2$ if it was at $x_1$ in the instant $t_1$. A nice but
somewhat heuristic way to obtain this amplitude is to use a path
integral. The main contribution to it comes from the value that the
action takes on the classical path connecting $(x_1,t_1)$ to
$(x_2,t_2)$. However, we also have to consider the contributions given
by other paths, especially those ``close'' to the classical
trajectory. It is clear now that the amplitude will depend both on the
class of paths used in the definition of the integral and the specific
form of the action that has to be evaluated on these. Notice that it
is possible to have different Lagrangians leading to the same
equations of motion. Furthermore, these Lagrangians do not necessarily
differ from each other in total derivative or divergence terms
(an example of this phenomenon in the context of general relativity is
provided by the self-dual action
\cite{Jacobson:1988y,Samuel:1987td} and the Holst action
\cite{Holst:1995pc}). One expects that a modification -- either in the
class of allowed paths and/or in the action -- will generically change
physical amplitudes.

A natural way to think about a quantum symmetry reduction of a model
(again in the heuristic setting provided by path integrals) would
consist in first restricting ourselves to computing probability
amplitudes between symmetric configurations and then considering only
a restricted class of paths in the path integral: precisely those that
are, themselves, symmetric. This would have two important
effects. First, the value of the probability amplitude will
generically differ from the one obtained by considering unrestricted
trajectories connecting the two symmetric initial and final
configurations. This is expected because we are ignoring paths that
would be taken into account for the non-reduced system. Second, it
will be generally impossible to recover the amplitudes corresponding
to the full theory from the symmetry reduced ones because information
is unavoidably lost in the process of rejecting the non-symmetric
trajectories (which can be thought of as a projection,
see~\cite{Engle:2005ca} and also \cite{Torre:2008dv} for a more general
point of view). This is even more so because, in principle,
completely different mechanical or field systems may have the same
reduced sectors under a given symmetry.

Though it can be argued that we can learn very important lessons from
a quantum symmetry reduced model, and even get significant qualitative
information about the full quantized theory, it will be generally
impossible to recover exact results referring to the latter. This
would be so even if we restricted ourselves to computing transition
amplitudes between symmetric classical configurations. A nice
discussion on this issue appears in~\cite{Kuchar:1989tj}. There the
authors compare in a quantitative way the physical predictions derived
from two different symmetry reductions of general relativity such that
one of higher symmetry (the Taub model) is embedded in the other (the
mixmaster model). They do this by constructing appropriate inner
products and comparing the probabilistic interpretations of wave
functions in both models. Their conclusion is in a way expected: the
respective behaviors are different. This result sends an important
warning signal: one should not blindly extrapolate the results
obtained from the study of symmetry reductions. On the other hand it
does not exclude that in physically relevant situations one can
actually obtain interesting and meaningful predictions from the study
of the quantization of symmetry reductions.

Finally it is also important to disentangle this last issue from the
different one of understanding to what extent the processes of
symmetry reduction and quantization commute. To see this consider a
certain classical field theory derived from an action principle and a
symmetry reduction thereof obtained by restricting the action to
symmetric configurations (this procedure will be consistent if the
principle of symmetric criticality holds as we will discuss in the
next section). One can consider at this point the quantization of the
classical reduced model by using as the starting point the reduced
action. Supposing that one has a consistent quantization of the full
theory, one can try to see if it is possible to recover the results
obtained by first reducing and then quantizing by a suitable
restriction -- requiring a correct and consistent implementation of
the symmetry requirement -- of the fully quantized model. This has
been done in detail for the specific example of a rotationally
symmetric Klein--Gordon field in~\cite{Engle:2005ca}. The main result
of this paper is that it is indeed possible to show that using a
suitable ``quantum symmetry reduction'' both procedures give the same
result; i.e.\ in a definite sense reduction and quantization commute.

Giving a general prescription guaranteeing the commutation of both
procedures on general grounds would be certainly a remarkable result,
specially if applicable to instances such as loop quantum
gravity. This is so because many details of the quantization of full
general relativity in this framework are still missing. It would be
very interesting indeed to know what LQG would say about concrete
symmetry reductions of full quantum gravity that could conceivably be
obtained by considering the comparatively simpler problem of
loop-quantizing the corresponding reduced classical gravity
model. This notwithstanding one should not forget what we said
above. Even if this can be effectively done we would not learn the
answer to the problem of computing the amplitudes predicted by LQG for
transitions between symmetric configurations. This implies that the
results derived in symmetry reduced implementations of the full LQG
program such as loop quantum cosmology, no matter how suggestive they
are, cannot be extrapolated to completely trustable predictions of
full quantum gravity.

%================================================================
\newpage

\section{Midisuperspaces: Classical Aspects}
\label{sect4}

A rough classification of the symmetry reductions of general
relativity can be made by considering the dimension of the isometry
group of the metric. In many cases this is equivalent to classifying
the spacetime metrics for the midisuperspace model in question
according to the number of Killing vector fields that they
have. Though this is a sensible approach, especially when the Killing
fields commute, this is not always the most natural way to describe
all the interesting symmetry reductions, in particular, when spherical
symmetry is present.

\subsection{One-Killing vector reductions}

Let us start by considering the simplest example of symmetry reduction
corresponding to one-dimensional spatial isometry groups. In practice
one considers $\mathbf{R}$ or $U(1)$ and, hence, the spacetime metrics
are required to have a single Killing vector field. These models are
interesting because they retain important features of full general
relativity, in particular diffeomorphism invariance, an infinite
number of physical degrees of freedom and a non-linear character.

The local aspects of the one-Killing vector reductions were first
considered by Geroch~\cite{Geroch:1970nt}. In that paper he developed
a method to dimensionally reduce gravity by defining a way to
``project'' 3+1 dimensional geometric objects to the 2+1 dimensional
space of orbits of the Killing field (required to have a non-vanishing
norm). Here the world ``local'' refers to the fact that some
topological aspects are sidestepped in a first look; however, if the
quotient itself is well behaved (for example, it is Haussdorff) the
projection is globally defined and has a clear geometrical meaning.
The most important result of this work was to show that the reduced
system can be interpreted as 2+1 general relativity coupled to certain
matter fields with a concrete geometrical meaning: the norm and
twist of the four dimensional Killing vector field (a scalar and a
one-form field respectively). This link between one-Killing vector
reductions and 2+1 dimensional gravity theories opened the door to
quantum treatments relying on techniques specially tailored for lower
dimensional models. The Geroch method can be adapted to treat symmetry
reductions. For example it allows to write the four dimensional scalar
curvature as a curvature on the 2+1 dimensional orbit manifold plus
some extra terms involving the norm of the Killing. This is very
useful to write the 3+1 dimensional action as a 2+1 dimensional one.

The global aspects and the Hamiltonian formalism (for vacuum GR) have
been discussed by Moncrief in the case when the symmetry group is
$U(1)$ with compact Cauchy surfaces~\cite{Moncrief}. The spatial
slices in this case can be taken to be $U(1)$ bundles (or rather
$S^1$) over the sphere (though the analysis can be extended to
arbitrary surfaces). The discussion presented in~\cite{Moncrief} is
relevant to study some of the compact Gowdy models, in particular
those with the $S^2\times S^1$ and $S^3$ spatial topologies, though it
is possible to employ other approaches that rely on the Geroch
reduction as discussed in~\cite{Barbero:2007vg}. The non-compact case
with asymptotically flat two-geometries (in the sense relevant in 2+1
gravity developed in~\cite{Ashtekar:1993ds}) has been studied by
Varadarajan~\cite{Varadarajan:1995hw}.

\subsection{Two-Killing vector reductions}

The next natural step is to consider two-dimensional spacelike
isometry groups. A local approach that parallels the one given for the
one-Killing vector reduction was developed by Geroch
in~\cite{Geroch:1972yt} for the abelian case (corresponding to
commuting Killing fields). The global aspects for these models, in the
case of considering commutative, connected and two dimensional
isometry groups with effective and proper action, have been discussed
in detail in~\cite{Berger:1994sm,Chrusciel1990100}. In the following
we will simply refer to them as \textit{two-Killing vector reductions}
despite the fact that they do not correspond to the most general
situation.

The classification of the smooth, effective, proper actions by
isometries of a commutative, connected, two-dimensional Lie group on a
connected, smooth 3-manifold has been studied in the literature both
for the compact~\cite{Mostert1,Mostert2} and
non-compact~\cite{Berger:1994sm} topologies. Of all the possible
choices of group action and spatial topology only two have been
considered with sufficient detail from the quantum point of view:

\begin{itemize}

\item The Einstein--Rosen cylindrical waves, with isometry group
  $\mathbf{R}\times U(1)$ and spatial topology $\mathbf{R}^3$.

\item The Gowdy models, whose isometry group is $U(1)\times U(1)$ and
  the spatial topologies are $T^3:=S^1\times S^1\times S^1$,
  $S^2\times S^1$, $S^3$, and the lens spaces.

\end{itemize}

In the restricted case when the group orbits are hypersurface
orthogonal we have the so called \textit{polarized models}
(also known as linearly polarized models). Otherwise we have general
polarization (or non-polarized) models. Historically
two-Killing vector reductions were introduced to explore some concrete
problems; in particular, the original motivation by Einstein and
Rosen~\cite{Einstein:1937qu, Beck} was to use cylindrical symmetry
as a simplifying assumption to explore the existence of gravitational
waves (see, however,~\cite{Kennefick}). Gowdy considered the
$U(1)\times U(1)$ model as a first step to study inhomogeneous
cosmologies~\cite{Gowdy:1971jh,Gowdy:1973mu}. As we are mainly
interested in the quantization of midisuperspaces we discuss next only
those classical issues of direct use in the quantum treatment of these
models, in particular their Hamiltonian description.

The first Hamiltonian analysis of the Einstein--Rosen waves was
carried out by Kucha\v{r} in the early seventies~\cite{Kuchar:1971xm},
however a complete treatment incorporating the appropriate surface
terms had to wait until the nineties~\cite{Ashtekar:1996bb}. A work
that was very influential to get the final Hamiltonian formulation
was~\cite{Ashtekar:1993ds}. In this reference, Ashtekar and
Varadarajan studied the Hamiltonian formalism of asymptotically flat
2+1 dimensional general relativity coupled to matter fields satisfying
positive energy conditions. They found that, due to the peculiarities
associated with the definition of asymptotic flatness in the 2+1
setting, the Hamiltonian of those systems is always bounded both from
below \textit{and from above}. This is a very important result for the
class of midisuperspaces whose 2+1 description can be performed in terms
of a non-compact Cauchy surfaces. As we have pointed out before, this is
the case for certain one-Killing symmetry reductions and especially
for Einstein--Rosen waves~\cite{Ashtekar:1996bb}. Before we proceed
further we want to point out that the asymptotic analysis mentioned
above has been extended to discuss the structure of the null infinity
in the 2+1 description of one-Killing symmetry reductions of 3+1 GR
in~\cite{Ashtekar:1996cd} and the behavior of Einstein--Rosen waves
at null infinity has been studied in detail in~\cite{Ashtekar:1996cm}.

In the case of the Einstein--Rosen waves the two Killing vector fields
correspond to translations and rotations. The translational Killing
field has non-vanishing norm whereas the rotational one vanishes at
the symmetry axis. The main steps to arrive at the Hamiltonian
formulation of the Einstein--Rosen waves can be summarized as follows:

\begin{itemize}

\item Start from the Einstein--Hilbert action with the appropriate
  surface terms depending on the extrinsic curvature at the boundary.

\item Use then the translational Killing vector to perform a Geroch
  reduction and obtain a 2+1 dimensional action written in terms of
  the 2+1 dimensional metric in the space of orbits and a scalar field
  (the norm of the Killing).

\item Get the Hamiltonian from this 2+1 dimensional action in the
  standard way by using an axially symmetric foliation adapted to the
  Minkowskian observers with proper time $t$ at infinity. In this
  process it is necessary to introduce and take into account the
  fall-off conditions for the fields corresponding to the relevant
  asymptotics~\cite{Ashtekar:1996bb} and the fact that the action
  depends on a Minkowskian fiducial asymptotic metric.

\item Finally use a gauge fixing condition to select a single point in
  each gauge orbit.

\end{itemize}

After all these steps the Hamiltonian is obtained in closed form as a
function of the free Hamiltonian $H_0$ for an axially symmetric
massless scalar field evolving in an auxiliary, 2+1 dimensional
Minkowskian background
\begin{equation}
\label{ham}
H=2\Big(1-\exp(-H_0/2)\Big).
\end{equation}
The fact that this Hamiltonian is a function of a free one can be used
to study the exact classical dynamics of this system. It is very
important to point out that the Minkowskian metric (induced by the
metric in the asymptotic region) plays an auxiliary role. Although the
Hamilton equations are non-linear it is possible to achieve a
remarkable simplification by introducing a redefined time variable of
the form $T:=\exp(-H_0/2) t$ where $t$ is the asymptotic inertial
time. When this is done the scalar field that encodes the
gravitational degrees of freedom of the model must simply satisfy the
free field equation for an axisymmetric field. This fact allows us to
quantize the model by using a Fock space. In particular the exact
unitary evolution operator can be written in closed form and, hence,
closed form expressions can be written for many interesting
physical objects such as two-point functions.

We want to mention that some generalizations of the Einstein--Rosen
waves to a class of cylindrical spacetimes endowed with angular
momentum have been considered by Manojlovi\'c and Mena
in~\cite{Manojlovic:2000nw}, where the authors found boundary
conditions that ensure that the resulting reduced system has a
consistent Hamiltonian dynamics. This work lead
Mena~\cite{MenaMarugan:2000gk} to consider a definition of cylindrical
space-times that is less restrictive than the usually employed in the
literature. The key idea is to remove the symmetry axis from the spacetime
and, as a consequence, relaxing the regularity conditions in the fields
when they approach this spacetime boundary. Little is presently known
about the quantization of these systems so we will not consider them
further.

The Hamiltonian formalism for the Gowdy models has been developed by
many authors both in vacuum~\cite{MenaMarugan:1997us,Pierri:2000ri}
and coupled to scalar fields for all the possible
topologies~\cite{Barbero:2007vg}. The Hamiltonian analysis for the
$T^3$ topology can be seen
in~\cite{MenaMarugan:1997us,Pierri:2000ri}. An interesting technical
point that is relevant here concerns gauge fixing. In this model the
natural gauge fixing condition gives rise to a so called
\textit{deparametrization} because the fixing is not
complete. Actually after this partial gauge fixing two first class
constraints remain. One of them is linear in the momentum canonically
conjugate to a variable that can be interpreted as time. As a consequence
of this it is possible to reinterpret the system as one described by
a time dependent Hamiltonian. The other constraint remains as a
condition on physically acceptable configurations. A thorough discussion
on this issues can be found in~\cite{Barbero:2007vg}. This last paper
carefully extends the Hamiltonian analysis to the other possible spatial
topologies. In particular it discusses the constraints that must be
included in the Hamiltonian formulation to take into account the regularity
conditions on the metric in the symmetry axis for the $S^2\times S^1$
and $S^3$ topologies. The main difficulty that arises now is the vanishing
of some of the (rotational) Killing vector fields at some spacetime
points. In the case of the $S^2\times S^1$ topology only one of the
two Killings vanish so the problem can be dealt with by using the
non-vanishing Killing to perform a first Geroch reduction and
carefully deal with the second by imposing suitable regularity
conditions for the fields in the 2+1 dimensional formulation. The
$S^3$ case is harder to solve because \textit{both} Killing fields
vanish somewhere. Nevertheless the problem can be successfully tackled
by excising the axes from the spacetime manifold and imposing suitable
regularity conditions on the fields when they approach the artificial
boundary thus introduced. An interesting feature that shows up in the
Hamiltonian analysis for the $S^2\times S^1$ and $S^3$ topologies is
the presence of the so called \textit{polar constraints} induced by
the regularity conditions. The final description for these topologies
is somehow similar to the one corresponding to the $T^3$ case, in
particular the fact that the dynamics of the system can be described
with a time dependent Hamiltonian that clearly shows how the initial
and final singularities appear. The main difference is the absence of
the extra constraint present in the $T^3$ model. This is a consequence
of the details of the deparametrization process.

Other classical aspects related to Gowdy models coupled with matter
(concerning integrability or the obtention of exact solutions) have
been covered in detail in~\cite{Charach:1978vf, Charach:1979xh,
  Charach:1980xu, Carmeli:1981nk, Carmeli:1983ny}; in particular, the
identification of a complete set of Dirac observables for the
Einstein--Rosen and the Gowdy $T^3$ midi-superspace was obtained
in~\cite{Torre:1991gn, Torre:2005cn, Kouletsis:2003hj} (and
in~\cite{Husain:1994dc} by using Ashtekar variables). The relation of
two-Killing reductions and sigma and chiral models have been
considered by many authors~\cite{Ashtekar:1997zf, Husain:1995fd,
  McGuigan:1990nd}. We want to mention also an interesting paper by
Romano and Torre~\cite{Romano:1995ep} where they investigate the
possibility of developing an internal time formalism for this type of
symmetry reduction. They also show there that the Hamiltonian of these
models corresponds to that of a parametrized field theory of axially
symmetric harmonic maps from a 3-dimensional flat spacetime to a
2-dimensional manifold endowed with a constant negative curvature
metric (though in the compact cases the presence of constraints must
be taken into account).

\subsection{Spherical symmetry}

Spherically symmetric systems in general relativity are another type
of midisuperspace models (in a sense ``the other type'') that have
received a lot of attention. They are enjoying a second youth these
days as very useful test beds for LQG. A 3+1 spacetime $(M,g)$ is
called spherically symmetric if its isometry group contains a subgroup
isomorphic to $SO(3)$ and the orbits of this subgroup are 2-spheres
such that the metric $g$ induces Riemannian metrics on them that are
proportional to the unit round metric on $S^2$. Notice that in the
standard definition of spherically symmetry the spacetime manifold is
taken to be diffeomorphic to $\mathbf{R}\times \Sigma$, where the
Cauchy hypersurface $\Sigma$ is $\mathbf{R}\times S^2$ (notice,
however, that this is not the only possibility~\cite{Clarke, Siegl,
  Szenthe}). In this case the $SO(3)$ symmetry group acts without
fixed points (there is no center of symmetry). The spherically
symmetric metric on the Cauchy slices $\mathbf{R}\times S^2$ is given
by $\Lambda^2(t,r)dr^2+R^2(t,r)d\Omega^2$ (where $d\Omega^2$ denotes
the metric of the unit 2-sphere). This metric depends only on two
functions $\Lambda(t,r)$ and $R(t,r)$. The radial coordinate $r\in
\mathbf{R}$ is such that the $r\rightarrow \pm \infty$ limits
correspond to the two different spatial infinities of the full
Schwarzschild extension and $t$ denotes a ``time'' coordinate.

The first attempt to study spherically symmetric models in general
relativity from the Hamiltonian ADM point of view goes back to the
paper by Berger, Chitre, Moncrief and Nutku~\cite{Berger:1972pg}. Here
the authors considered vacuum gravity and also coupled to other fields
such as massless scalars. The problem with this approach, as pointed
out by Unruh in~\cite{Unruh:1976db}, was that they did not reproduce
the field equations. The cause for this was identified also by Unruh:
a boundary term needed to guarantee the differentiability of the
Hamiltonian was missing in the original derivation. It must be pointed
our that the paper~\cite{Berger:1972pg} predates the classic one by
Regge and Teitelboim~\cite{Regge:1974zd} where the role of surface
terms in the correct definition of the Hamiltonian framework for
general relativity is discussed in detail. We want to mention also
that~\cite{Berger:1972pg} was the starting point of an interesting
series of articles by H\'aj\'{\i}\v{c}ek on Hawking
radiation~\cite{Hajicek:1984je, Hajicek:1984mz, Hajicek:1984nb}.

The spacetime slicings chosen in the first studies of spherically
symmetric models covered only the static regions of the extended
Schwarzschild spacetime (the Kruskal extension). This means that, in
practice, they only considered the Schwarzschild geometry outside the
event horizon. This problem was tackled by Lund~\cite{Lund} who used a
different type of slicing that, however, did not cover the whole
Kruskal spacetime with a single foliation. An interesting issue that
was explored in this paper had to do with the general problem of
finding a canonical transformation leading to constraints that could
give rise to a generalized Schr\"{o}dinger representation (as was done
by Kucha\v{r} in the case of cylindrical
symmetry~\cite{Kuchar:1971xm}). One of the conclusions of this
analysis was that this was, in fact, impossible; i.e.\ there is no
``time variable'' such that the constraints are linear in its
canonically conjugate momentum. This negative conclusion was, however,
sidestepped by Kucha\v{r} in~\cite{Kuchar:1994zk} by cleverly using a less
restrictive setting in which he considered foliations going form one
of the asymptotic regions of the full Kruskal extension to the
other. This paper by Kucha\v{r}~\cite{Kuchar:1994zk} is in a sense the
culmination of the continued effort to understand the quantization of
Schwarzschild black holes in the traditional geometrodynamical
setting. It must be said, however, that it was predated by the
analysis performed by Thiemann and Kastrup~\cite{Kastrup:1993br,
  Thiemann:1992jj} on the canonical treatment of Schwarzschild black
holes in the Ashtekar formalism. In~\cite{Kastrup:1993br} the authors
found a pair of canonical variables that coordinatize the reduced
phase space for spherically symmetric black holes consisting of two
phase space variables $M$ and $T$ where $M$ is the black hole mass and
$T$ is its conjugate variable that can be interpreted as ``time''
(more precisely the difference of two time variables associated with
the two spatial asymptotic regions of an eternal black hole). This
description of the reduced phase space precisely coincides with the
ones given by Kucha\v{r}~\cite{Kuchar:1971xm}. We want to mention also
that an interesting extension of of Kucha\v{r}'s work appears
in~\cite{Varadarajan:2000jy}. In this paper Varadarajan gave a
non-singular transformation from the usual ADM phase space variables
on the phase space of Schwarzschild black holes to a new set of
variables corresponding to Kruskal coordinates. In this way it was
possible to avoid the singularities appearing in the canonical
transformations used by Kucha\v{r}.

The Hamiltonian formulations obtained by these methods provide a
precise geometrical description of the reduced phase for vacuum
spherically symmetric general relativity. In particular an exact
parametrization of the reduced phase space is achieved. At this point
it is just fitting to quote Kucha\v{r}~\cite{Kuchar:1994zk}:

\bigskip

``\textit{Primordial black holes, despite all the care needed for
  their proper canonical treatment, are dynamically trivial.}''

\bigskip

A possible way to have spherically symmetric gravitational models with
local degrees of freedom and avoid this apparent triviality consists
in coupling matter to gravity. It must be said, nevertheless, that for
some types of matter couplings the reduced phase space of spherically
symmetric systems is finite dimensional. This is so, for example, in
the case of adding infinitesimally spherical thin shells. The
Hamiltonian analysis of the massive and the null-dust shell cases has
been extensively studied in the literature~\cite{Friedman:1997fu,
  Berezin:1997fn, Louko:1997wc, Hajicek:1998hd, Hajicek:2000mh}. The
presence of additional null shells has been also
analyzed~\cite{Hajicek:2001ky, Hajicek:2001kz}.

It is perhaps more surprising to realize that this finite-dimensional
character is also a property of spherically symmetric Einstein--Maxwell
spacetimes with a negative cosmological constant, for which the gauge
symmetries exclude spherically symmetric local degrees of freedom in
the reduced phase space. In this case canonical transformations of the
Kucha\v{r} type can be used~\cite{Louko:1996dw} to obtain the reduced
phase space Hamiltonian formulation for the system. Once matter in the
form of massless scalar fields is coupled to gravity, the reduced
system is a (1+1)-dimensional field theory and some of the techniques
developed by Kucha\v{r} cannot be applied. In particular, Romano has
shown~\cite{Romano:1995ff} that the coupled Einstein-Klein Gordon
system does not have a suitable extrinsic time variable. As we
mentioned above, the Hamiltonian formulation for the gravity-scalar
field model was clarified by Unruh in~\cite{Unruh:1976db}. Recently,
some simplifications have been obtained by using flat slice
Painlev\'e--Gullstrand coordinates~\cite{Husain:2005gx}. Other types of
matter that can be coupled to gravity giving rise to infinite
dimensional reduced phase spaces are those including collapsing dust
clouds~\cite{Vaz:2001bd,Vaz:2001mb}.

To end this section, we should mention that another interesting way to
gain insights into the quantization of more realistic gravity models,
such as the collapse of spherically symmetric matter in 3+1
dimensions, is to consider two-dimensional dilaton gravity as in the
Callan, Giddings, Harvey and Strominger (CGHS)
model~\cite{Callan:1992rs} -- and similar ones that admit a phase space
description close to the 3+1 spherically symmetric spacetimes. These
systems are usually exactly solvable (both classically and quantum
mechanically) and hence can be used to study the consequences of
quantizing gravity and matter. From a technical point of view these
models are close to spherical symmetry because they can be treated by
using the same type of canonical transformations introduced by
Kucha\v{r} in~\cite{Kuchar:1994zk}. They lead to descriptions that are
quite close to the ones obtained for the vacuum Schwarzschild
case~\cite{Varadarajan:1995jj, Bose:1995gy, Gegenberg:2009ny}.

%================================================================
\newpage

\section{Midisuperspaces: Quantization}
\label{sect5}

\subsection{Quantization of one-Killing vector reductions}

The quantization of this class of midisuperspaces has been considered
by some authors but it is fair to say at this point that only a very
partial knowledge has been achieved. A preliminary analysis within the
loop quantum gravity framework (with complex variables) was carried
out in~\cite{Husain:1989mp}. The fact that these models can be
interpreted as 2+1 dimensional gravity coupled to some matter fields
suggests that a quantization of this lower dimensional system
essentially solves the problem for one-Killing vector
reductions. In fact, some claims about the perturbative
renormalizability of 2+1 gravity coupled to scalar fields have
appeared in the literature~\cite{Bonacina:1992fw} although, in our
opinion, it is rather unclear how these results can be
extrapolated to symmetric gravity in 3+1 dimensions. The reason for
this scepticism is the non-trivial structure of the Hamiltonian for
these systems when the right asymptotic behavior is incorporated
(taking into account that, as shown by Ashtekar and Varadarajan
in~\cite{Ashtekar:1993ds}, the Hamiltonian that corresponds to the
generator of time translations at spatial infinity is actually bounded
from above). In any case, the physical consequences of these
perturbative analyses have not been explored.

\subsection{Quantization of two-Killing vector reductions}

\subsubsection{Quantization of Einstein--Rosen waves}

The first historic attempt to canonically quantize (vacuum)
Einstein--Rosen waves goes back to the pioneering paper by
Kucha\v{r}~\cite{Kuchar:1971xm}. There he carefully studied
cylindrical metrics for the polarized case and derived the
Hamiltonian formulation for the system. The author used the full four
dimensional picture in a very effective way in order to develop a
suitable Hamiltonian formalism, in particular, the canonical
transformations leading to a convenient coordinatization of the
reduced phase space. One of the key achievements of the paper was the
identification of a phase space function that could play the role of a
time variable for ER-waves. This provides an extrinsic time
representation similar to the one used in the spherically symmetric
case~\cite{Kuchar:1994zk}. By defining an appropriate canonical
transformation it is possible to turn this time into a canonical
variable and make it part of a new set of canonical coordinates. In
terms of them the action functional takes the particularly simple
form of the parameterized formalism for an axially symmetric scalar
field evolving in a (fictitious) Minkowskian background. An
interesting comment is that the canonical transformation mentioned
above mixes configuration and momentum variables in such a way that
the original configuration space is traded for a rather different one
which is not a subset of a space of metrics. The main problem
with~\cite{Kuchar:1971xm} was that it did not take into account the
necessary boundary terms needed to render the variational problem
well-defined. The results of the derivation given by Kucha\v{r} can be
obtained in a more systematic and straightforward way by using the
principle of symmetric criticality~\cite{Romano:1995ep}, substituting
the form of the cylindrically symmetric metrics corresponding to
polarized Einstein--Rosen waves in the Einstein--Hilbert
action and getting the Hamiltonian formulation from there.

The Dirac quantization of the Einstein--Rosen waves that Kucha\v{r}
gives is interesting from a pedagogical and intuitive point of view
but arguably quite formal. The main consistency issues related, for
example, with the path independence with respect with the foliations
interpolating between two given ones are formally taken into account
as well as the definition of the scalar product of Schr\"{o}dinger
picture functionals. However, no attention is payed to the subtle
functional analytic and measure theoretic issues that come up. This
problem has been addressed in~\cite{Cho:2006zz,Corichi:2002ir}. Cho
and Varadarajan~\cite{Cho:2006zz} have studied the relationship between the
Schr\"{o}dinger and Fock representations and considered several issues
related to the unitary implementability of the evolution of the free
axisymmetric scalar field in a Minkowskian background. In particular
they have discussed the existence of unitary transformations on the
Fock space implementing the evolution between two axisymmetric, but
otherwise arbitrary, Cauchy slices of the auxiliary flat spacetime in
such a way that their infinitesimal version gives the functional
Schr\"odinger equations obtained by Kucha\v{r}. The analysis is based
on work by Torre and Varadarajan~\cite{Torre:1997zs,Torre:1998eq} on
the evolution of free scalar fields between arbitrary Cauchy
surfaces. In this respect it is interesting to remark the different
behaviors in the 1+1 dimensional case and the higher dimensional ones.
It is also important to point out that polarized Einstein--Rosen
waves are remarkably close to this type of model. The
main result of~\cite{Cho:2006zz} is that in the half parameterized
case, when the radial coordinate is not changed, the dynamics can be
unitarily implemented as a consequence of Shale's
theorem~\cite{Shale}. However, if no condition is imposed on the
radial coordinate -- thus allowing for the possibility of having
radial diffeomorphisms -- the quantum counterpart of these
transformations cannot be unitarily implemented because it is not
given by a Hilbert--Schmidt operator. The unitary equivalence of the
Schr\"{o}dinger and Fock quantizations for free scalar fields has been
studied in a slightly more general setting by Corichi et
al.\ in~\cite{Corichi:2002ir}. The authors of this paper take into
account several functional issues that are relevant for a rigorous
treatment of the rather subtle issues that crop up in the quantization
of free field theories defined in arbitrary globally hyperbolic
spacetimes.

A different approach to the quantization of Einstein--Rosen waves by
standard quantum field theory methods consists in performing a
complete gauge fixing and studying the reduced space of the
model. This has been done by Ashtekar and
Pierri~\cite{Ashtekar:1996bb}. The quantization of the system is
performed after paying special attention to the asymptotic conditions
relevant in the 2+1 dimensional case (and, consequently, to the
necessary boundary terms in the gravitational action). Specifically,
some of the peculiarities of this system come from the fact that, due
to the translational part of the isometry group, the (non-trivial)
Einstein--Rosen solutions cannot be asymptotically flat in four
dimensions (or alternatively, the 2+1 dimensional metric does not
approach a Minkowski metric at spatial infinity because a deficit
angle is allowed). The numerical value of the Hamiltonian of this
system, when the constraints are satisfied, is given by an expression
originating in the surface term of the action for Einstein--Rosen
waves. As mentioned before this has the form given by
Equation~(\ref{ham}) and is a function of $H_0$, the free Hamiltonian
corresponding to an axisymmetric massless scalar field in 2+1
dimensions evolving in a Minkowskian background metric. This allows us
to use a Fock space to quantize the system. In fact, as the
Hamiltonian is just a function of a free one it is possible to obtain
the \textit{exact} quantum evolution operator from the one
corresponding to the free auxiliary model and use it to obtain close
form expressions for many objects of physical interest such as
field commutators and $n$-point functions. It is important to point
out that only after the gauge fixing is performed the Hilbert space of
states takes this form. This means that other quantizations or gauge
fixings describing the gravitational degrees of freedom in a different
way, could lead to a different form for the quantized model.
It is important to notice that Ashtekar and Pierri do not
perform the canonical transformation used by Kucha\v{r} to introduce
the extrinsic time representation but directly fix the gauge in the
original canonical formulation. By using this scheme
Varadarajan~\cite{Varadarajan:1999aa} has studied the mathematical
properties of the regularized quantum counterpart of the energy of the
scalar field in a spherical region of finite radio of 2+1 flat
spacetime. In particular he has given a proof of the fact that this
regularized operator is densely defined and discussed its possible
self-adjoint extensions (this is only a symmetric operator).

In the aftermath of~\cite{Ashtekar:1996bb} several papers have used
the quantization presented there to derive physical consequences of
the quantization of Einstein--Rosen waves. Among the most influential
is~\cite{Ashtekar:1996yk}. In this paper Ashtekar considers 2+1
gravity coupled to a Maxwell field -- with the additional condition of
axisymmetry -- as a solvable toy model to discuss quantum gravity
issues. It is important to point out that in 2+1 dimensions the
Maxwell field can be interpreted as a free massless scalar, with
the usual coupling to gravity, and dynamics given by the wave
equation. This system is equivalent to the symmetry reduction of 3+1
dimensional gravity provided by Einstein--Rosen waves so the results
on the paper are relevant to study the quantum physics of this
midisuperspace model. The main result of~\cite{Ashtekar:1996yk} is the
somewhat surprising appearance of \textit{large quantum gravity
  effects} in the system. The presence of these has some importance
because, if we trust this quantization, it points out to the spurious
character of many classical solutions to the model as they cannot
appear in the classical limit. One key element here is the form of the
Hamiltonian, given by Equation~(\ref{ham}) and the fact that it is a
non-linear function of a free Hamiltonian. The large quantum gravity
effects manifest themselves in the expectation values of the metric
components (that are functions of the scalar field that describes the
degrees of freedom). Specifically these do not correspond to the
classical values in some limits, in particular for high
frequencies. Furthermore the metric fluctuations in this regime are
very large even for states that are peaked around a classical
configuration of the scalar field and grow with the number of quanta
(``photons'') of the scalar field. A reflection of these large
quantum gravitational effects is also manifest in the behavior of the
field commutators, especially at the symmetry axis, as shown
in~\cite{BarberoG.:2004uc}.

An extension of these results to four dimensions has been carried out
in the paper by Angulo and Mena~\cite{Angulo:2000ad}. They do this
by expressing the four-dimensional metric of the Einstein--Rosen waves
in terms of the Maxwell scalar and the 2+1 dimensional metric. It is
important to mention at this point that the scalar field enters the
four-metric in a highly non-linear way. The most important result
of~\cite{Angulo:2000ad} is the actual verification of the possibility
to extend the conclusions reached by Ashtekar in the 2+1 dimensional
setting to four dimensions (far from the symmetry axis). However the
authors argue that to reach an acceptable classical description in the
asymptotic region in four dimensions it is not mandatory to require
-- as in three -- that the number of quanta of the Maxwell scalar
field be small. Also the behavior on the symmetry axis is
interesting because the relative uncertainties of the metric become
very large there. This casts some doubts on the appropriateness of the
classical regularity conditions usually introduced at the axis.

There are other papers where the physical consequences of the Fock
quantization of Einstein--Rosen waves given in~\cite{Ashtekar:1996bb}
have been considered. In~\cite{BarberoG.:2003ye, BarberoG.:2004uc,
  BarberoG.:2004uv, BarberoG.:2005ge, BarberoG.:2006gd,
  BarberoG.:2008ei} several physical issues have been discussed in
some detail, in particular microcausality, $n$-point functions,
2-point functions, matter couplings and coherent
states. Microcausality in Poincar\'e-invariant models, formulated with
the help of a background Minkowskian metric, reflects itself in the
vanishing of the commutator of the field operators at spatially
separated spacetime points. The fact that Einstein--Rosen waves can
be quantized offers the possibility of quantitatively testing
some of the expected features of quantum gravity such as the smearing
of light cones due to quantum fluctuations of the metric. The presence
of this effect was suggested already in the original paper by Ashtekar
and Pierri~\cite{Ashtekar:1996bb}. In~\cite{BarberoG.:2003ye} the
explicit form of the commutator was obtained and a direct numerical
analysis showed the expected smearing effect and clear indications
about how the classical limit is reached when the effective
gravitational constant of the model goes to zero. A quantitative
understanding of this phenomenon has been given
in~\cite{BarberoG.:2004uc, BarberoG.:2004uv} by performing an
asymptotic analysis of the integrals that define the field
propagators. An interesting result coming from these analyses is the
different way in which the classical limit is reached on and outside
the symmetry axis. Outside the symmetry axis the large scale regime is
reached in a rather smooth way but on the axis large quantum
gravitational effects persist even at macroscopic scales. This is
probably another manifestation of the kind of effects described by
Ashtekar in~\cite{Ashtekar:1996yk}.

A way to incorporate quantum test particles -- that could further help
explore the quantized geometry of the model -- is to couple matter
fields to gravity and use their quanta as probes. This is very
difficult for generic matter fields but can be remarkably achieved for
massless scalars keeping both the classical and quantum resolubility
of the system. This was done in~\cite{BarberoG.:2005ge,
  BarberoG.:2006gd} -- though the classical integrability was
understood by several preceding authors, in particular Lapedes,
Charach, Malin, Feinstein, Carmeli and
Chandrasekhar~\cite{Lapedes:1977sz, Charach:1978vf, Charach:1980xu,
  Chandrasekhar:1986jp}. It is also fair to mention at this point that
the effective decoupling of the gravitational and matter scalar modes
(in the flat space picture) that is the key ingredient in the Fock
quantization of the Einstein--Rosen waves coupled to massless scalar
fields presented in~\cite{BarberoG.:2005ge} was discussed in
essentially the same form by Lapedes in~\cite{Lapedes:1977sz} though
he treated the quantization in the heuristic way customary at the
time.

In~\cite{BarberoG.:2006gd} Newton--Wigner localized states were used
to build actual position space wave functions for the massless quanta
in order to study how they evolve in full interaction with the
quantized geometry. The resulting picture shows, in a convincing way,
that the quantum particles in their motion define approximate
trajectories that follow the light cones given by the
microcausality analysis. Also the study of 2-point functions (extended
to $n$-point functions in~\cite{BarberoG.:2008ei}) gives a consistent
picture when they are interpreted as approximate propagation
amplitudes; in particular the persistence of large quantum gravity
effects in the symmetry axis is confirmed. Finally the issue of
obtaining coherent states for the Einstein--Rosen waves has been
considered in~\cite{BarberoG.:2008ei}.

The results described above refer to polarized ER-waves. It
is natural then to consider the full non-polarized case. An
interesting set of papers by Korotkin, Nicolai and
Samtleben~\cite{Nicolai:1996pd, Korotkin:1994au, Korotkin:1995zj,
  Korotkin:1996vi, Korotkin:1997ps} explores a family of systems
consisting of two-dimensional general relativity coupled to non-linear
sigma models. These generalize the symmetry reductions from 3+1
dimensions that we are considering in this section, in particular
the Einstein--Rosen waves and treat genuine midisuperspace
models with non-linear interactions and an infinite number of degrees
of freedom. A unified treatment of them appears
in~\cite{Nicolai:1996pd} where a number of issues related to their
classical integrability and quantization are discussed. By assuming
the presence of two commuting Killing symmetries it is possible to
make a first simplification of the functional form of the metrics
for this system by restricting the number of coordinates on
which the fields depend to just two. In this way they effectively
correspond to 2-dimensional non-linear sigma models coupled to a
dilaton and gravity. A key observation is that the resulting models
are integrable and their solutions can be obtained from an auxiliary
linear system of equations. In fact, the matter dynamics can be
derived from the compatibility conditions for this linear system. The
Hamiltonian formulation can be written in terms of two constraints
generating translations along the light cone that become partial
partial differential equations when quantized. These quantum
constraints are precisely the Knizhnik-Zamolodchikov equations
that play a fundamental role in conformal field theory. Their
solutions, as discussed in~\cite{Nicolai:1996pd}, provide concrete
physical states of the quantized theory. Some of the specific models
collectively considered in this paper are individually studied in a
series of works by the authors~\cite{Korotkin:1994au, Korotkin:1995zj,
  Korotkin:1996vi}. The limits of this line of work are related to the
impossibility of solving the coset constraints for the non-compact
spaces [such as $SL(2,\mathbf{R})/SO(2)$] by using discrete unitary
representations [of $SL(2,\mathbf{R})$ in the previous example].

We want to mention here that the paper~\cite{Korotkin:1997ps} is of
particular interest because Korotkin and Samtleben extend the Fock
quantization techniques used in~\cite{Ashtekar:1996bb} to the much
harder non-polarized case where the non-linearity of the model
shows up in full strength. In this case the Einstein field equations
can be written as the so called Ernst equation for the unimodular
metric on the Killing orbits and an integral expression for the
conformal factor of the metric in terms of the solutions to this Ernst
equation. The quantization is achieved by finding a complete set of
quantum observables and a representation of them in a Fock Hilbert
space. This is done by introducing a new set of classical
observables. These generalize another set of variables that can be
defined in the polarized case in terms of the positive and
negative frequency modes that appear in the Fourier decomposition of
the axisymmetric scalar field that encodes the gravitational degrees
of freedom. The Poisson algebra of the new observables is quadratic
(i.e. the Poisson bracket of two basic observables can be written as
a linear combination of products of two of them); this introduces
some complications in the treatment due to the necessity to deal with
operator ordering issues after quantization. However the solution to
this problem is known in the theory of integrable
systems~\cite{Maison:1978es}. The final step of
the process consists in finding a representation of the quantum
algebra in a Fock Hilbert space. The availability of this
representation opens the possibility of computing expectation values
of important operators, in particular certain components of the
metric. In principle this can be used to derive physics from the
non-polarized ER-waves although the non-local character of these new
observables introduced by the authors and, in particular, their lack
of an explicit spacetime dependence makes it difficult to make
contact with other results, specially those related to the study of
microcausality in the polarized case~\cite{BarberoG.:2003ye}.
As a final comment on the approach by Korotkin, Nicolai and Samtleben
we want to mention the possibility of mapping the non-polarized
cylindrical models to free theories as discussed in~\cite{Cruz:1998di}.

To end this section we want to comment that the quantization of
Einstein--Rosen waves has also been considered from other points of
view such as perturbative methods and LQG inspired techniques. The use
of perturbative techniques in particular is a very natural path to
follow because it offers the possibility of comparing the results with
those obtained by non-perturbative methods and ultimately try to get
an additional understanding of the causes leading to the perturbative
non-renormalizability of general
relativity. In~\cite{Niedermaier:2002eq} a thorough study of the
quantization of two-Killing vector reductions is carried out. The main
goal of this paper was to find out if the perturbative quantization of
Einstein--Rosen waves is consistent with the asymptotic safety
scenario of Weinberg~\cite{Niedermaier:2006wt}. In fact the main
result of~\cite{Niedermaier:2002eq} is to show that two-Killing
symmetry reductions of general relativity are asymptotically
safe. However, this result is not straightforward because
two-Killing vector reductions are not renormalizable in the standard
sense (beyond one loop). Despite this the model can be declared to be
renormalizable if the space of Lagrangians is expanded by allowing
conformal factors that are functions of the so called radion field
(the determinant of the pull-back of the metric to the integral
manifolds of the Killing vector fields). In fact, when this is done
the renormalization flow has a unique ultraviolet stable fixed point
where the trace anomaly vanishes~\cite{Niedermaier:2002eq}. A similar
result has been obtained in the case of non-polarized Einstein--Rosen
waves in~\cite{BarberoG.:2003pz} by using a path integral approach and
the algorithm developed by Osborn in~\cite{Osborn:1987au} to deal with
position dependent sigma models. These papers provide and alternative
point of view from the esentially non-perturbative methods of
Korotkin, Nicolai and Samtleben. A comparison of physical predictions
in both approaches would be most interesting. Several other papers can
be found in the literature that consider different aspects of some
two-Killing vector reductions from the perturbative point of view; in
particular~\cite{BarberoG.:2004bi}

The loop quantization of the Einstein--Rosen midi-superspace is an
interesting open problem that deserves some comments. In their seminal
paper about Fock quantization of cylindrically symmetric
spacetimes~\cite{Ashtekar:1996bb}, Ashtekar and Pierri computed the
holonomies around those loops that are integral curves of the
rotational Killing vector and showed that their traces are functions
of the energy (in a box of finite radius) of the scalar field encoding
the reduced degrees of freedom. In particular, in the large radius
limit, those traces reduce to a simple function of the total energy of
the system. Hence, as they point out in the paper, the question if
those traces are well-defined operators in the quantum theory reduces
to the question of whether the operator corresponding the energy of a
scalar field in a box can be satisfactorily regulated
(see~\cite{Varadarajan:1999aa}). In any case, the dynamical
issues of the polymeric quantization of the scalar field (including
the classical limit and the relation with standard quantizations) need
to be analyzed in detail.

As usual in loop quantum gravity, one follows a two-step route to
quantization by first constructing the kinematical Hilbert space of
the theory and then defining the Hamiltonian constraint (and, for ER
waves, also the Hamiltonian) of the model. In this last step, the
geometric operators (in particular the volume operator) are thought to
play a relevant role in the rigorous definition of the Hamiltonian
constrain operator. In his Living Review~\cite{Bojowald:2006da}
Bojowald discusses the kinematical Hilbert for space for ER-waves and
also certain properties of the volume operator. He pays special
attention to the differences of the cylindrically symmetric sector and
the homogeneous cosmologies. There are also other papers, much more
qualitative in nature, by Neville~\cite{Neville:2005ve,
  Neville:2005vf} where the construction of a kinematical Hilbert
space for the loop quantization of cylindrically symmetric spacetimes
and planar waves is sketched.

\subsubsection{Quantization of Gowdy models}

Gowdy cosmological models~\cite{Gowdy:1971jh,Gowdy:1973mu} are
described by time oriented, globally hyperbolic, vacuum spacetimes
which can be constructed from the evolution of $U(1)\times
U(1)$-invariant Cauchy data defined on a 3-dimensional closed (compact
without boundary) hypersurface $\Sigma$. The action of the
$U(1)\times U(1)$ group of spatial isometries is assumed to be
effective and the topology of $\Sigma$ is restricted to be $T^3$,
$S^2\times S^1$, $S^3$, or the lens spaces. These spacetimes describe
inhomogeneous cosmologies with initial and/or final
singularities and represent gravitational waves propagating in a closed
universe. For the $T^3$ topology only a singularity is present whereas
in the case of the $S^2\times S^1$, $S^3$ topologies there are both
initial and final singularities. The quantization of these
models has been considered only in the polarized case (for
which both Killing vector fields are hypersurface orthogonal).

The Gowdy $T^3$ model describes an inhomogeneous cosmology with one
singularity (that can be thought of as initial or final). Charles
Misner was the first researcher to recognize its relevance as a test
bed for quantum cosmology~\cite{Misner:1972js,
  Misner:1973zz}. Pioneering work on its quantization was carried out
in the 1970s and 1980s by him and
Berger~\cite{Berger:1975kn, Berger:1974, Berger:1984iya}. Actually it
is fair to say that many of the ideas that have been used to this day
in the attempts to achieve a rigorous quantization for this system
actually originate in these works. In particular:

\begin{itemize}

\item The Hamiltonian analysis for this type of models, including the
  identification of the extra constraint present for the $T^3$
  topology~\cite{Misner:1973zz, Berger:1974}.

\item The introduction of the deparametrization customarily used to
  achieve a (partial) gauge fixing~\cite{Misner:1973zz}.

\item The identification of a suitable real Fourier expansion that
  decouples the modes appearing in the
  Hamiltonian~\cite{Misner:1973zz}.

\item The identification of the time-dependent field redefinitions
  that have been used in recent works to find a unitary implementation
  of the quantum dynamics~\cite{Berger:1974} (though these were used
  by Berger in a different context to study the WBK regime).

\item The role of the harmonic oscillator with time-dependent
  frequency in the dynamics of the Gowdy
  models~\cite{Misner:1973zz, Berger:1974}.

\end{itemize}

The approach to quantization followed in these papers consisted in a
rather formal treatment in which the Hilbert space was taken to be the
infinite tensor product of the (countably) infinite Hilbert spaces
associated with each of the oscillator modes appearing in the Fourier
transformation of the fields. The use of an infinite tensor product of
Hilbert spaces is problematic, as emphasized by
Wald~\cite{Wald:1995yp}, because this Hilbert space is non-separable
and the representation of the canonical commutation relations is
reducible. Misner and Berger have discussed a number of issues
related to graviton pair creation in Gowdy universes and the
semiclassical limit. A nice and intuitive picture developed in these
papers is the idea that their quantum dynamics can be interpreted as
scattering in superspace.

An interesting problem that received significant attention even at
this early stage was the issue of the singularity resolution. Berger
approached it by working in a semiclassical approximation where it
was possible to describe the gravitational radiation by means of an
effective energy momentum tensor depending on some of the metric
components. By using this framework it was argued that the classical
singularity was replaced by a bounce. This very same question was
considered by Husain in~\cite{Husain:1987am} where he used the same
kind of Hilbert space as Berger but followed a different approach
consisting in quantizing the Kretschmann invariant (the square of the
Riemann tensor) after finding an appropriate operator ordering. This
was done by imposing the sensible requirement that the expectation
values of the Kretsschmann operator in coherent states equal their
classical values sufficiently far from the singularity. The main
result of~\cite{Husain:1987am} was that, at variance with the findings
of Berger, the classical singularity persisted in the quantized
model. It is interesting to mention that this quantum version of the
Kretschmann invariant for Gowdy $T^3$, was also used by Husain to
explore a conjecture by Penrose pointing a relation between the Weyl
curvature tensor and gravitational entropy~\cite{Husain:1988jf}.

In the mid nineties, the work on the quantization of the
Einstein--Rosen waves developed by Ashtekar and
Pierri~\cite{Ashtekar:1996bb} led naturally to the revision of the
quantization of other two-Killing midisuperspace models, and in
particular the Gowdy $T^3$ cosmologies. This was done, among other
reasons, to open up the possibility of using this type of symmetry
reductions as toy models for LQG. Since then the system has been
considered not only within the traditional geometrodynamical approach
but also in the Ashtekar variables framework. The Hamiltonian
formalism of the Gowdy $T^3$ model, with a detailed analysis of the
gauge fixing procedure, was studied in terms of (complex) Ashtekar
variables by Mena in~\cite{MenaMarugan:1997us} (see
also~\cite{Husain:1989qq}). In that paper, the quantization of the
reduced model was also sketched, however the first attempt to study the
Fock quantization of the polarized $T^3$ Gowdy model (in the
geometrodynamical setting) appears in~\cite{Pierri:2000ri}. In this
paper, Pierri used one of the $U(1)$ subgroups of the isometry group
to perform a dimensional reduction and represent the model as 2+1
gravity coupled to a massless scalar field. She showed that the
reduced phase space could be identified with the one corresponding to
a $U(1)$ symmetric, massless scalar field propagating in a 2+1
background geometry and satisfying a quadratic constraint. By using
this description of the reduced phase space, she proposed a Fock
quantization that relied on a quantization of the modes of the free
field propagating in the 2+1 background geometry. The quadratic
constraint was imposed \textit{\`a la Dirac}. The main drawback of
this approach, as later pointed out by Corichi, Cortez and Quevedo
in~\cite{Corichi:2002vy}, was that the quantum dynamics of the free
scalar field used in Pierri's quantization does not admit an unitary
implementation. It is important to realize that this type of behavior
is not a specific pathology of the Gowdy models but actually an
expected (and somehow generic) feature of quantum field
theories~\cite{Helfer:1996my}. The results of~\cite{Corichi:2002vy}
were confirmed and extended by Torre~\cite{Torre:2002xt} who was able
to show that that, even after restricting the quantum dynamics to the
physical Hilbert space obtained by imposing the constraint present in
the model {\textit{\`{a} la Dirac}}, the quantum evolution is not
given by a unitary operator.

This important problem was tackled and solved in a satisfactory way in
a series of papers by Corichi, Cortez, Mena, and
Velhinho~\cite{Corichi:2005jb, Corichi:2006xi, Corichi:2006zv,
  Corichi:2007ht, Cortez:2005th, Cortez:2007hr, Cortez:2008hk,
  Cortez:2009zv}. These authors have shown that it is actually
possible to have unitary dynamics if one redefines the basic scalar
field in the description of the Gowdy $T^3$
model~\cite{Corichi:2005jb, Corichi:2006xi} by introducing an
appropriate time-dependent factor (inspired by a similar field
redefinition used by Berger in~\cite{Berger:1974}). An additional
important uniqueness result appearing in these papers is that, up to
unitary equivalence, this is the only way in which the dynamics can be
unitarily implemented in this reduced phase space quantization of the
system.

The quantum description of the $S^1\times S^2$ and $S^3$ Gowdy models
in terms of a Fock quantization of their reduced phase spaces can be
found in~\cite{BarberoG.:2007rd} (the details of the Hamiltonian
formulation for these topologies were studied
in~\cite{Barbero:2007vg}). In those cases, the reduced phase spaces
can be identified with the ones corresponding to $U(1)$-symmetric
massless scalar fields. The problem of unitarily implementing the
quantum dynamics is present also for these topologies but, as in the
$T^3$ case, the quantum dynamics can be implemented in an unitary way
if the scalar fields are suitably redefined~\cite{BarberoG.:2007rd, Vergel:2008eb, Cortez:2009zv}.

The description of the reduced phase space of Gowdy models in terms of
massless scalar fields has been used to explore the quantum
Schr\"odinger representation of the system in terms of
square-integrable functions on a space of distributional fields with a
Gaussian probability measure~\cite{Torre:2006kt, Corichi:2007ht,
  Vergel:2008eb, Vergel:2009st}. This representation is, in this case,
unitarily equivalent to the Fock one but in some situations it is
actually more convenient due to the availability of a spacetime interpretation.

We want to end this section by commenting that Gowdy models have been
used as a test bed to other approaches to quantum gravity.
In~\cite{Gambini:2005pg}, the $T^3$ model has been
studied within the ``consistent discretizations'' approach
(though the paper mostly deals with classical issues aimed at
the problem of showing that the discretizations reasonably reproduce
the expected classical results, in particular the preservation of
constraints). The analysis presented in this paper is relevant because
the difficulties to determine the lapse and shift suggest, according
to the authors, that the quantization of the non-polarized case will
have to rely on numerical methods.

The loop quantization of the non-polarized Gowdy $T^3$ model in terms
of complex Ashtekar variables was considered for the first time by
Husain and Smolin in~\cite{Husain:1989qq}. In this work, Husain and
Smolin found a loop representation of the unconstrained algebra of
observables and gave sense to the (regulated) constrains in this
representation. They also constructed a large and non-trivial sector
of the physical state space and identified the algebra of physical
operators on the state space. The loop quantization of the
polarized Gowdy $T^3$ model has been studied recently in terms of real
Ashtekar variables by Barnerjee and Date~\cite{Banerjee:2007kh,
  Banerjee:2007ki} where the authors recast the model in terms of
real $SU(2)$ connections as a first step towards quantization and also
discuss the gauge fixing procedure. A preliminary description of the
kinematical Hilbert space for the polarized Gowdy $T^3$ model and some
issues related to the volume operator are given
in~\cite{Banerjee:2007ki}.

The previous papers do not discuss the quantum resolution of the
classical singularity, a natural question to consider after the
success of LQC in the study of this problem in the simpler setting
provided by homogeneous models. This very important issue has been
recently considered by using a combination of loop and Fock
quantizations in~\cite{MartinBenito:2008ej,Brizuela:2009nk}. In
reference~\cite{MartinBenito:2008ej} the authors use the formulation
of the theory as 2+1 gravity coupled to a massless scalar field (with
a residual $U(1)$ symmetry). By introducing the usual Fourier mode
decomposition of the solutions in terms of the relevant angle
variable, the authors define a hybrid quantization consisting in a
polymeric quantization for the homogeneous mode (angle-independent)
and a Fock quantization for the inhomogeneous (angle-dependent)
ones. The main result of the article is that the
singularity is resolved. The follow-up study appearing
in~\cite{Brizuela:2009nk} further considers the quantum dynamics of
the polarized Gowdy $T^3$ model, in particular the description of the
initial Big Bang singularity that appears to be replaced by a Big
Bounce as in the popular LQC models.

\subsubsection{Other related models}

Two-Killing vector reductions of general relativity, in the case when
the Killing vectors are hypersurface orthogonal, can be classified
according to some properties of the gradient of the determinant of the
restriction of the metric to the group orbits (the area function). The
familiar cases of the Einstein--Rosen waves and the Gowdy cosmologies
correspond to spacelike and timelike gradients respectively whereas
\textit{plane waves} correspond to the null case. The
geometrodynamical approach to the quantization of plane waves
midisuperspaces was considered in~\cite{MenaMarugan:1998uf} where the
authors study both the polarized and non-polarized cases. They show
that the reduced phase space models have vanishing Hamiltonians in the
coordinates adopted for their description. In the case of polarized
plane waves, the reduced phase space can be described by an
infinite set of annihilation and creation like variables (that are
classical constants of motion) and therefore it is possible to
quantize the system by finding a Fock representation for these
variables. In this respect the model is quite similar to the Gowdy
cosmologies and Einstein--Rosen waves that can also be quantized by
using Fock space techniques. The results of this paper have been used
in~\cite{MenaMarugan:1999et} to study the appearance of large quantum
effects in the system (similar to the ones described by Ashtekar for
Einstein--Rosen waves in~\cite{Ashtekar:1996yk}). The plane wave case
in rather interesting in this respect because of the focusing of light
cones characteristic of this system. Important quantum gravity effects
are expected precisely at the places where this focusing takes
place. By introducing suitable coherent states the authors show that
the expectation value of a regularized metric operator coincides with
a classical plane wave solution whereas the fluctuations of the metric
become large \textit{precisely} in the vicinity of the regions when
focussing of light cones occurs (and this happens for every coherent
state). These papers nicely complement and expand in a rigorous
language the preliminary analysis carried out in~\cite{Neville:1997jy,
  Borissov:1994uc} in terms of (complex) Ashtekar variables. A
complete discussion based in the modern approach to connection dynamics
and symmetry reductions in this framework would be interesting indeed.

Finally we want to mention another cosmological midisuperspace model,
due to Schmidt~\cite{Schmidt}, that has some interesting features. The
spacetime of this case has the topology $\mathbf{R}^2\times T^2$,
i.e.\ it is the product of a plane and a torus, and the isometry group
is $U(1)\times U(1)$ with orbits given by tori. Schmidt cosmologies
have initial singularities that are similar to Gowdy
$T^3$. Beetle~\cite{Beetle:1998iu} has studied its Hamiltonian
formulation (in the polarized case) by choosing appropriate
asymptotic conditions for the fields in such a way that the resulting
reduced phase space is very similar to the one corresponding to the
Gowdy $T^3$ model. In fact the system can be described as a $U(1)$
symmetric massless scalar field evolving in a fixed time dependent 2+1
background (topologically $\mathbf{R}^2\times S^1$) and with no extra
constraints (at variance with the Gowdy $T^3$ case). The same
unitarity problems that show up in the quantum evolution of the Gowdy
models also appear here and can be solved again with a time dependent
canonical transformation~\cite{BarberoG.:2006zw}.

\subsection{Spherically symmetric midisuperspace}

Spherically symmetric reductions of general relativity have an obvious
appeal as they can serve both as interesting toy models to test
fundamental aspects of quantum general relativity and also describe
very interesting physical situations involving Schwarzschild black
holes. Spherical symmetry is peculiar in the sense that classically
the space of spherically symmetric solutions to vacuum general
relativity is parameterized by a single quantity -- the black hole mass
-- but general spherically symmetric metrics require the introduction
of \textit{functions} for their most general description that can be
taken to depend only in a radial
coordinate and a time variable. As a consequence, and although the
reduced phase space for vacuum spherical general relativity is
actually finite dimensional (this is essentially the content of
Birkhoff's theorem), the Hamiltonian analysis for this type of system
must be performed in the infinite dimensional phase space
corresponding to a field theory. This means that, in principle, there
is an important difference between a reduced phase space quantization
(or one involving gauge fixing) where the field theory aspects and the
issues related to diff-invariance are hidden, and the of use a Dirac
approach that will require the introduction of constraint operators to
select the physical states from those in the linear space of quantum
states of a proper field theory.

To further complicate matters (or actually making them more
interesting) some parts of the Schwarzschild solution can be
faithfully described with minisuperspace models. This is the case for
the interior Schwarzschild metric that can be isometrically mapped to
a Kantowski-Sachs model. As we will mention later, this fact has been
used to discuss issues related to the quantum resolution of black hole
singularities by adapting loop quantum cosmology methods. Though this
Living Review is devoted to midisuperspace models, for the sake of
completeness we deem it necessary to include a discussion of the
relevant results obtained in this minisuperspace setting, specially
those devoted to black hole singularity resolution, before dealing with
the midisuperspace aspects of spherical symmetry.

\subsubsection{Singularity resolution in a minisuperspace
  approximation: Black hole interior}

Early suggestions of singularity resolution coming from the use of LQG
inspired techniques go back to the works of Husain, Winkler and
Modesto. In~\cite{Modesto:2004xx} an exotic quantization developed
in~\cite{Husain:2003ry} was used to study singularity resolution in a
geometrodynamical description of spherical black holes. Later
in~\cite{Husain:2004yz}, the well known fact that the interior of a
Schwarzschild black hole is isometric to the Kantowski--Sachs
homogeneous cosmological model is mentioned. This idea, combined with
the exotic quantizations considered in~\cite{Husain:2004yz}, leads in
a natural way to the claim that black hole singularities are resolved
in this scheme. Though some attempts to derive this result within the
LQG framework by quantizing the Kantowski--Sachs model appear in the
literature~\cite{Modesto:2004wm, Modesto:2005zm}, the first complete
and rigorous account of it is due to Ashtekar and
Bojowald~\cite{Ashtekar:2005qt}. In this paper the authors import some
of the results obtained in the framework of loop quantum cosmology to
the study of black hole interiors and the issue of singularity
resolution. The quantization is performed by first describing the
system in real connection variables. This is important because the
degenerate triads play a significant role in the quantization process
-- in particular the phase space points corresponding to them do not
lie on a boundary of the constraint hypersurface. These singular
configurations can be naturally accommodated in the Ashtekar variables
framework but not in standard geometrodynamics. Then the loop
quantization of the Kantowski-Sachs model is carefully carried out to
completion. Some important results can be derived from it, in
particular it is possible to prove that the physical spacetime
corresponding to the interior of a black hole does not end at the
classical singularity but can be extended beyond it. On the other hand
the quantum fluctuations close to the singularity are such that a
classical description breaks down. The issues of matter coupling and
the extension of the quantization to the exterior region outside the
horizon were not considered in~\cite{Ashtekar:2005qt}. These problems
and the work related to them will be mentioned and described in an
upcoming Section~\ref{ssect541} devoted to the loop quantization of
spherically symmetric spacetimes.

Within a semiclassical approach the study of what is the kind of
physical object that replaces the classical singularity after
quantization has been attempted by several researchers by employing
different approximations. In~\cite{Bohmer:2007wi} the authors suggest
that the original quantization used in~\cite{Ashtekar:2005qt} may not
have the correct semiclassical limit (as it happens when an analogue
quantization is performed in loop quantum cosmology). The proposed way to
remedy this problem was to allow the parameter (playing the role of
the $\mu_0$ \textit{polymeric} parameter of LQC) appearing in the
construction of the Hamiltonian constraint operator to be a
\textit{function} of the triads (in the same way as $\mu_0$ is allowed
to depend of the scale factor in LQC) and work with an effective
Hamiltonian constraint. This generalizes and improves the results
of~\cite{Modesto:2006mx} (see also~\cite{Modesto:2008im} for other
different types of suggested improvements) where Modesto uses exactly
the approach of~\cite{Ashtekar:2005qt}. Though the results presented
in these works are not rigorous derivations within LQG, they support
the idea that the singularity is resolved and replaced
by either a wormhole type of solution or a ``swollen'' singularity
described by a spherical surface that is asymptotically approached but
never reached by infalling particles. Further refinements on the ideas
and approximations presented in these papers can be found
in~\cite{Chiou:2008nm} (where a fractal type of spacetime is generated
by the creation of a series of smaller and smaller black holes spawned
by quantum collapses and bounces).

More general examples have been considered in the minisuperspace
framework and using LQG inspired methods. For example, the quantum
collapse of an spherical dust cloud is described
in~\cite{Modesto:2006qh} both in the ADM and Ashtekar formulations and
some issues concerning black hole evaporation are discussed
in~\cite{Modesto:2009ve}. Additional references on the minisuperspace
treatment of black hole singularities
are~\cite{Modesto:2007wp, Peltola:2008pa, Peltola:2009jm}.

\subsubsection{Quantization of spherically symmetric midisuperspaces}

In this section, we will review the literature relevant to study the
geometrodynamical quantization of spherical midisuperspace models, an
upcoming section will describe the results that are being obtained
these days by using loop quantum gravity inspired methods. As a
general comment we would like to say that the level of mathematical
rigor used in the standard quantization of these systems is sufficient
in some cases but, as a rule, the attention payed to functional
analytic issues and other mathematical fine points does not reach the
level that is standard now in loop quantum gravity. This should not be
taken as a criticism but as a challenge to raise the mathematical
standards. In our opinion a rigorous treatment within the
geometrodynamical framework could be very useful in order to deepen
our understanding of quantum gravity (see~\cite{Alink} for some
interesting results in this respect).

A fundamental paper in the study of the classical and quantum behavior
of spherically symmetry reductions of general relativity is due to
Kucha\v{r}~\cite{Kuchar:1994zk}. This work provides several key
results for these systems within the geometrodynamical framework. For
example, it gives canonical coordinates leading to a very simple
description of the reduced phase space for an eternal black hole and
explains along the way how the transformation to this coordinate
system can be obtained from the knowledge of the extended
Schwarzschild solution (in Kruskal coordinates). This canonical
transformation is widely used to analyze the Hamiltonian dynamics of
other spherically symmetric gravitational systems coupled to matter as
will be commented later in this section. This work discusses the
quantization of the system in several possible schemes: the direct
reduced phase space quantization and the Dirac approach. Finally, it
proves the unitary equivalence of the resulting quantum models. This
paper is a basic reference where subtle but important issues are taken
into account, in particular those related to the asymptotic behavior
of the fields and the introduction of the necessary boundary terms in
the Einstein--Hilbert action. It must be said, nonetheless, that the
identification of the canonical pair of variables that describe the
reduced phase space for eternal black holes was first found by Kastrup
and Thiemann in their study of the very same system within the
Ashtekar variables approach~\cite{Thiemann:1992jj, Kastrup:1993br}.

The quantization of Schwarzshild black holes was also considered
around the same time in~\cite{Cavaglia:1994yc, Cavaglia:1995bb}. These
papers rely on the study of what the authors call the
$r$-dynamics. The reason why they have to use some non-standard
dynamics for the system to develop a Hamiltonian formalism is
that the parametrization of the spherical metrics that they employ
involves only functions of a \textit{single radial coordinate} $r$
(and do not involve an additional ``time'' variable $t$). This amounts
to introducing from the start the ansatz that the metrics are not only
spherical but also static. These papers rely also in the use of a
rather formal Wheeler--DeWitt approach to quantization. This means
that some well-known consistency problems are
present here. It must be said that, at the end of the day, some of the
results of~\cite{Cavaglia:1994yc, Cavaglia:1995bb} are compatible with
the ones discussed~\cite{Kuchar:1994zk} and~\cite{Kastrup:1993br} (in
particular the unavoidable identification of the black hole mass as
the configuration variable for the spherical black hole reduced phase
space) but in our opinion this approach has been superseded by the
treatment provided by Kucha\v{r}, Kastrup and Thiemann that considers
the full Kruskal spacetime for eternal black holes and discuses the
Hamiltonian framework and the identification of the reduced phase
space in that setting.

An interesting way to go beyond vacuum spherically symmetric gravity
is by coupling a single particle-like degree of freedom, which in this
case corresponds to an infinitely thin spherical matter shell. After
the system is quantized this shell degree of freedom can be
conceivably used as a quantum test particle allowing us to extract
interesting information about the (spherical) quantized geometry. The
paper by Berezin, Boyarsky and Neronov~\cite{Berezin:1997fn} studies
this system both form the classical and quantum points of view by
using the standard geometrodynamical approach. One of the key
ingredients is the use of the canonical tranformation introduced by
Kucha\v{r} in his famous paper on spherical symmetric
gravity~\cite{Kuchar:1994zk}. Although the quantization presented
in~\cite{Berezin:1997fn} is quite formal, some interesting features are
worth a comment, in particular the fact that the Schr\"{o}dinger
equation becomes a difference equation (a feature reminiscent of
results obtained with LQG methods). By using analyticity arguments the
authors identify the quantum numbers characterizing the system
(actually a pair of integer numbers that parameterize the mass
spectrum). This result, however, does not correspond to the one
proposed by Bekenstein and Mukhanov and derived by other authors in
different frameworks.

The collapse of a thin null dust shell has been
extensively studied by H\'aj\'{\i}\v{c}ek in the context of a
midisuperspace quantization~\cite{Hajicek:1998hd, Hajicek:2002ny,
  Hajicek:2000mh, Hajicek:2001ky, Hajicek:2001kz}. Classically this
system gives rise to black holes and, hence, its quantization may shed
light on the issue of singularity resolution. In this case the unitary
evolution of the wave packets representing the collapsing shell degree
of freedom suggest that the singularity is resolved because they
vanish at the place where the singularity is expected to be
(see~\cite{Hajicek:2002ny} and references therein).

A series of papers~\cite{Vaz:1998gm, Vaz:1999nt, Vaz:2000mq,
  Vaz:2000zb, Vaz:2001bd, Vaz:2001mb, Vaz:2002xb, Vaz:2009jj,
  Vaz:2009mu, Kiefer:2005tw, Kiefer:2007va} dealing with the
quantization of spherically symmetric models coupled to matter is due
to Kiefer, Louko, Vaz, L.~Witten and collaborators. A very nice
summary of some of these results appears in the book by
Kiefer~\cite{Kieferlibro}. We will describe them briefly in the
following. The first paper that we will consider
is~\cite{Vaz:1998gm}. Here the authors use the Kucha\v{r} canonical
quantization and the idea of introducing a preferred dust-time
variable (the Brown--Kucha\v{r} proposal~\cite{Brown:1994py}) to
quantize eternal black holes. By an appropriate selection of the lapse
function it is possible to make the proper dust time to coincide with
the proper time of asymptotic observers. This selects a particular
time variable that can be used to describe the system (furthermore the
total energy can be seen to be the ADM mass of the black hole). An
interesting result derived in this reference is the quantization law
for the black hole mass of the type $M_n=\sqrt{n}m_P$ (where $m_P$
denotes the Planck mass and $n$ a positive integer) for the normalized
solutions of the Wheeler--DeWitt equation with definite parity (notice
that this is precisely the mass spectrum proposed by Mukhanov and
Bekenstein~\cite{Bekenstein, Mukhanov:1986me}). It is interesting to
point out that the authors do not use a purely reduced phase space
quantization but a Dirac approach. This means that the wave functions
for the quantized model do not depend only on the mass (in the vacuum
case) but also on some embedding variables that can be used to
distinguish between the interior and the exterior of the black
hole. The matching conditions for the wave function in these two
regions give rise to the quantization of the mass. The fact that the
wave function of the system vanishes outside and is different from
zero inside can be interpreted by realizing that observers outside the
horizon should see an static situation. The results obtained by
Kastrup, Thieman~\cite{Thiemann:1992jj} and
Kucha\v{r}~\cite{Kuchar:1994zk} have been also used by Kastrup
in~\cite{Kastrup:1996pu} to get similar results about the black hole
mass quantization.

The methods introduced in~\cite{Vaz:1998gm} were used in the
midisuperspace approach for the computation of black hole entropy
discussed in~\cite{Vaz:1999nt}. Here the authors have shown that it is
actually possible to reproduce the Bekenstein--Hawking law for the
entropy as a function of the black hole area after introducing a
suitable microcanonical ensemble. In order to do this they first showed
that the Wheeler--DeWitt wave functional can be written as a direct
product of a finite number of harmonic oscillator states that can
themselves be thought of as coming from the quantization of a massless
scalar field propagating in a flat, 1+1 dimensional background. The
finite number of oscillators, that originates in the discretization
introduced on the spatial hypersurfaces is then estimated by
maximizing the density of states. This is proportional to the black
hole area and suggests that the degrees of freedom responsible for the
entropy reside very close to the horizon. A curious feature of the
derivation is the fact that, in order to get the right coefficient for
the Bekenstein--Hawking area law, an undetermined constant must be
fixed. This is reminiscent of a similar situation in the standard LQG
derivation~\cite{Ashtekar:1997yu} where the value of the Immirzi
parameter must be fixed to get the correct coefficient for the
entropy. In our opinion there are some arbitrary elements in the
construction, such as the introduction of a discretization, that make
the entropy computation quite indirect.

Partially successful extensions of the methods used
in~\cite{Vaz:1998gm, Vaz:1999nt} to the study of other types of black
holes appear in~\cite{Vaz:2000mq, Vaz:2000zb, Vaz:2001bd}. In the first
of these papers the authors consider charged black holes. By solving
the functional Schr\"{o}dinger equation it is possible to see that the
\textit{difference} of the areas of the outer and inner horizons is
quantized as integer multiples of a single area. This is similar to
the Bekenstein area quantization proposal but not exactly the same because
the areas themselves are not quantized and the entropy is proportional
to this \textit{difference} of areas. The second paper studies gravitational
collapse as described by the LeMa\^{\i}tre--Tolman--Bondi models of
spherical dust collapse and considers a Dirac quantization of this
midisuperspace model. The main technical tool is, again, a
generalization of the methods developed by Kucha\v{r}, in particular
the canonical transformations introduced in~\cite{Kuchar:1994zk}. In
addition, the authors use the dust as a way to introduce a (natural)
time variable following the Brown--Kucha\v{r} proposal. The physical
consequences of this quantization (for the so called marginally bound
models) are explored in~\cite{Vaz:2001mb, Vaz:2002xb, Kiefer:2005tw,
  Kiefer:2007va} where the authors describe Hawking radiation and show
that the Bekenstein area spectrum and the black hole entropy can be
understood in a model of collapsing shells of matter. In particular
the mass quantization appears as a consequence of matching the wave
function and its derivative at the horizon. This result is compatible
with the one mentioned above for the Schwarzschild black
hole~\cite{Vaz:1998gm}. Some subtle issues, involving regularization
of the Wheeler--DeWitt equation are sidestepped by using the so called
DeWitt regularization $\delta^2(0)=0$, but the whole framework is
quite attractive and provides a nice perspective on quantum aspects of
gravitational collapse. In~\cite{Vaz:2001mb} dust is modeled as a
system consisting of a number of spherical shells. The entropy of the
black holes formed after the collapse of these $N$ shells depends,
naturally, on $N$. For black holes in equilibrium the authors estimate
this number by maximizing the entropy with respect to $N$ (this is
similar to the result mentioned before for the Schwarzschild black
hole). As far as Hawking radiation is concerned~\cite{Vaz:2002xb} the
authors model this system by taking the dust collapse model as a
classical background and quantizing a massless scalar field by using
standard techniques of quantum field theory in curved spacetimes. They
separately consider the formation of a black hole or a naked singularity.
In the first case they find that Hawking radiation is emitted
whereas in the second one the breaking of the semiclassical
approximation precludes the authors from deriving meaningful
results. The treatment provided in this paper is only approximate
(scalar products are not exactly conserved) and semiclassical (the WBK
approximation is used), but the resulting picture is again quite
compelling.

Black holes in AdS backgrounds have been considered
in~\cite{Vaz:2009jj, Vaz:2009mu} (see also~\cite{Franzen:2009ev}). The
most striking result coming from the analysis provided in these papers
is the fact that different statistics (Boltzmann and Bose--Einstein)
must be used in order to recover the correct behavior of the entropy
from the quantization of area in the Schwarzschild and large
cosmological constant limits respectively. Finally, we want to mention
that the collapse of null dust clouds has been partially discussed
in~\cite{Vaz:2001bd}.

A number of papers by Louko and collaborators~\cite{Louko:1994tv,
  Louko:1996dw, Louko:1996jd} study the Hamiltonian thermodynamics of
several types of black holes, in particular of the Schwarzschild,
Reissner--Nordstr\"om--anti-de~Sitter types, and black holes in
Lovelock gravity. The idea proposed in~\cite{Louko:1994tv} is to
consider a black hole inside a box and use appropriate boundary
conditions to fix the temperature. The black hole thermodynamics can
now be described by a canonical ensemble and standard statistical
physics methods can be used to compute the entropy. In particular the
authors provide a Lorentzian quantum theory and obtain from it a
thermodynamical partition function as the trace of the time evolution
operator analytically continued to imaginary time. From this partition
function it is possible to see that the heat capacity is positive and
the canonical ensemble thermodynamically stable. One of the remarkable
results presented in~\cite{Louko:1994tv} is that the partition
function thus obtained is the same as the one previously found by
standard Euclidean path integral methods~\cite{York:1986it,
  Braden:1987ad}.

Canonical transformations inspired in the one introduced by Kucha\v{r}
in~\cite{Kuchar:1994zk} play a central role in all these
analyses. Also the methods developed in~\cite{Kuchar:1994zk} have been
extended to study the thermodynamics of spherically symmetric
Einstein--Maxwell spacetimes with a negative cosmological
constant~\cite{Louko:1996dw} and spherically symmetric spacetimes
contained in a one parameter family of five-dimensional Lovelock
models~\cite{Louko:1996jd}. In both cases the canonical transformation
is used to find the reduced Hamiltonian describing these systems. The
most important conclusion of these works is that the Bekenstein--Hawking
entropy is recovered whenever the partition function is dominated by a
Euclidean black hole solution. In the Lovelock
case~\cite{Louko:1996jd} the results suggest that the thermodynamics
of five-dimensional Einstein gravity is rather robust with regard to
the the introduction of Lovelock terms. Another paper where the
Kucha\v{r} canonical transformation is used is~\cite{Kiefer:1998rr},
where the authors consider extremal black holes and how their
quantization can be obtained as a limit of non-extremal ones. The
obtention of the Bekenstein area quantization in this setting (for
Schwarzschild and Reissner--Nordstr\"om black holes) is described
in~\cite{Louko:1996md, Makela:1997rx}.

We end with a comment about the quantization of spherically symmetric
general relativity coupled to massless scalar fields. This has been
considered by Husain and Winkler in~\cite{Husain:2004yy}. In this
paper the authors study this problem in the geometrodynamical setting
by using Painlev\'e--Gullstrand coordinates (that are specially
suitable for this system). They use the non-standard quantization
described in~\cite{Husain:2004yz} that displays some of the features
of LQG and suggest that the black hole singularity is resolved. A
definition of \textit{quantum black hole} is proposed in the
paper. The key idea is to use operator analogues of the classical null
expansions and trapping conditions. As the authors emphasize, their
proposal can be used in dynamical situations (at variance with
isolated horizons~\cite{Ashtekar:2004cn}). The construction of
semiclassical states in this context and their use is further analyzed
in~\cite{Husain:2009vx}.

\subsubsection{Loop quantization of spherically symmetric models}
\label{ssect541}

Historically the first author to consider the treatment of spherically
symmetric gravity within the Ashtekar formalism was Bengtsson in
reference~\cite{Bengtsson:1989pp}. There he started to develop the
formalism needed to describe spherically symmetric complex $SU(2)$
connections and densitized triads and used it to discuss some
classical aspects related to the role of degenerate metrics in the
Ashtekar formulation and its connection with
Yang--Mills theories. The quantization of spherically symmetric
midisuperspace models written in terms of Ashtekar variables was
undertaken rather early in the development of LQG, by Kastrup and
Thiemann~\cite{Thiemann:1992jj}. At the time the mathematical setting
of loop quantum gravity was at an early stage of development and,
hence, the quantization that the authors carried out was not based in
the proper type of Hilbert space and made use of the old complex
formulation. Fortunately the reality conditions could be handled in
the spherically symmetric case. A key simplification is due to the
fact that the constraints can be written as expressions at most linear
in momenta. The resulting quantized model is essentially equivalent to
the geometrodynamical one due to Kucha\v{r}~\cite{Kuchar:1994zk}. An
interesting point that Kastrup and Thiemann discuss
in~\cite{Kastrup:1993br} is related to the physically acceptable range
of black hole masses (that somehow define the configuration space of
the model, at least in the reduced phase space formulation) and how
this should be taken into account when representing the algebra of
basic quantum operators. The (unitary) equivalence of the reduced and
Dirac quantizations for this system -- also found and discussed by
Kucha\v{r}~\cite{Kuchar:1994zk}-- can be proved once the right
ordering is found for the operators representing the constraints of
the system.

The modern approach to spherical symmetry reductions in loop quantum
gravity starts with~\cite{Bojowald:2004af}, where Bojowald carefully
introduced the necessary mathematical background to consider the
quantization of spherically symmetric models. In this paper Bojowald
constructs the kinematical framework for spherically symmetric quantum
gravity by using the full loop quantum gravity formalism; in
particular he shows how the states and basic operators (holonomies and
fluxes) can be derived from those in loop quantum gravity. An
important result of~\cite{Bojowald:2004af} is the realization of the
fact that significant simplifications take place that make these
symmetry reductions tractable. As expected they are midway between the
full theory and the homogeneous models that have been considered in
loop quantum cosmology. A very useful feature of the spherically
symmetric case, is the commutativity of the flux variables (thus
allowing for a flux representation in addition to the connection
representation that can be used as in the case of homogeneous
models). This work particularizes the general framework developed
in~\cite{Bojowald:1999eh} to study the quantum symmetry reduction of
diff-invariant theories of connections based on the isolation of
suitable symmetric states in the full 3+1 dimensional theory and the
subsequent restriction to this subspace (defining \textit{quantum
  symmetry reductions}).

The study of the volume operator for spherically symmetric reductions
was carried out in~\cite{Bojowald:2004ag} where its basic properties
were derived. In particular the volume operator was diagonalized and its
spectrum explicitly obtained. An important property of the
eigenfunctions is that they are not eigenstates of the flux operator
(and, in fact, have a complicated dependence on the connection). The
fact that on the volume eigenstates the holonomy operators have a
complicated dependence makes it quite difficult to study the
Hamiltonian constraint because it contains commutators of the volume
with holonomies. These are difficult to compute because the volume
eigenstates are not eigenstates of the triads (upon which the
holonomies act in a simple way). Nonetheless, an explicit construction
of the Hamiltonian constraint in the spherically symmetric case --
that makes use of non-standard variables that mix the connection and
the extrinsic curvature -- has been given in~\cite{Bojowald:2005cb}. These
new variables cannot be generalized but are specially tailored for the
spherically symmetric case. The main consequence of using them is the
simplification of the volume operator that they provide. This comes
about because the volume and flux eigenstates coincide for the new
variables. Notice, however, that the hard problem of finding the
kernel of the Hamiltonian constraint still has to be solved.

An obvious application of the formalism developed in these papers is
the study of singularity resolution for Schwarzschild black holes in
LQG. As shown in~\cite{Bojowald:2005ah} the structure of the quantized
Hamiltonian constraint for spherical symmetry reductions may allow us
to understand the disappearance of spacelike singularities. This issue
has also been considered for other spherically symmetric models such
as the Lema\^{i}tre--Tolman--Bondi collapse of an inhomogeneous dust
cloud~\cite{Bojowald:2008ja}. The (approximate) numerical analysis
carried out in this paper shows a slow-down of the collapse and
suggests that the curvature of naked singularities formed in
gravitational collapse can be weakened by loop quantum effects. This
is in agreement with the behavior expected in loop quantum gravity
where effective repulsive forces of a quantum origin usually make the
singularities tamer.

The main problem of these spherically symmetric approach followed by
Bojowald and collaborators -- as emphasized by Campiglia, Gambini and
Pullin in~\cite{Campiglia:2007pr} -- is related to the difficulties in
finding a particularization of the construction proposed by Thiemann
for the Hamiltonian constraint with the appropriate constraint algebra
in the diff-invariant space of states. This has led the authors
of~\cite{Campiglia:2007pr} to explore a different approach to the
quantization of spherically symmetric models in loop quantum
gravity~\cite{Campiglia:2007pr, Campiglia:2007pb, Gambini:2008dy,
  Gambini:2008ea, Gambini:2009ie} based on a partial gauge fixing of
the diffeomorphism invariance.

The quantization of the exterior Schwarzschild geometry has been
carried out in~\cite{Campiglia:2007pr} where the asymptotic behavior
of the fields relevant in this case was carefully considered. This
corrected a problem in~\cite{Bojowald:2005cb} related to the fall-off
of some connection components. The gauge fixing is performed in such a
way that the Gauss law is kept so that the reduced system has two sets
of constraints -- the Gauss law and the Hamiltonian constraint -- with
a non-trivial gauge algebra. Two approaches are then explored, in the
first of them the standard Dirac method is used after abelianizing the
constraints. The second is inspired by the fact that generically one
does not expect this abelianization to be possible. This led the
authors to use uniform discretizations~\cite{DiBartolo:2004cg,
  DiBartolo:2004dn} to deal with the general, non-abelianized
constraints. Although the study of the exterior region of a black hole
gives no information about singularity resolution, according to the
authors there are hints of singularity removal because the discrete
equations of the model are similar to those appearing in loop quantum
cosmology. The result of the loop quantum gravity quantization of the
exterior of a spherically symmetric black hole is in agreement with
the one obtained by Kucha\v{r} in terms of the usual geometrodynamical
variables; in particular the number and interpretation of the quantum
degrees of freedom (the mass of the spacetime) are the same in both
approaches. This means that the quantum dynamics is trivial: Wave
functions depend only on the mass and do not evolve.

After studying the exterior problem, the interior problem for a
Schwarzschild black hole was considered by these authors
in~\cite{Campiglia:2007pb} (in
a minisuperspace model similar to the one by Ashtekar and
Bojowald~\cite{Ashtekar:2005qt}). Here a suitable gauge fixing leads
to a description in terms of a Kantowski--Sachs metric. In this case it
is possible to describe the exact quantum evolution as a
semi-classical one with quantum corrections. The model is quantized in
the connection representation and behaves as a loop quantum cosmology
model where a certain type of bounce replaces the cosmological
singularity. When the quantum solution evolves past the singularity it
approaches another regime that behaves again in an approximate
classical way.

The complete black hole spacetime has been considered
in~\cite{Gambini:2008dy}. A key ingredient here is the choice of a
gauge fixing such that the radial component of the triad is a function
selected in such a way that in the limit when the ``polymerization''
parameter $\mu$ goes to zero one recovers the Schwarzschild metric in
Kruskal coordinates. The authors give a classical metric that
represents some of the features of the semiclassical limit for this
spherical black hole system where the singularity is effectively
avoided. The suggested picture consists of an eternal black hole
continued to another spacetime region with a Cauchy horizon. Far away
from the singularity the spacetime resembles the standard Schwarzshild
solution.

All these papers deal with partially gauge fixed Hamiltonian
systems. The issue of the residual diff-invariance in spherically
symmetric models quantized with LQG techniques is discussed
in~\cite{Gambini:2008ea}. It is shown there that it is possible
to reconstruct spacetime diffeomorphisms in terms of evolving
constants of motion (Dirac observables on the physical Hilbert
space), but some memory of the polymerization introduced by loop
variables remains because only a
subset of spacetime diffeomorphisms are effectively
implemented. According to Gambini and Pullin this is a reflection of
the fact that sub-Planck scales should behave differently from the
macroscopic ones.

Finally it is worth pointing out that the system consisting of
spherical gravity coupled to a massless scalar field has been
discussed in~\cite{Gambini:2009ie}. The same system was also
considered by Husain and Winkler in~\cite{Husain:2004yz} in the
context of geometrodynamical variables. In~\cite{Gambini:2009ie} the
authors use the uniform discretization technique to deal with the
thorny problem of working with a Lie algebra of constraints with
structure functions. Specifically they consider a discrete version of
the master constraint and use a variational method to minimize it. It
is important to understand that one should find the kernel of
this master constraint. The fact that this is not achieved in the
present model is interpreted by the authors as a hint that there is no
quantum continuum limit. Nevertheless the theory provides a good
approximation for general relativity for small values of the lattice
separation introduced in the discretization. The lowest eigenvalue of
the master constraint corresponds to a state with a natural physical
interpretation, i.e.\ the tensor product of the Fock vacuum for the
scalar field and a gaussian state centered around a flat spacetime
for the gravitational part. The authors argue that it is impossible
that loop quantum gravity regulates the short distance behavior of
this model in the gauge that they use. This leads them to conclude
that one should face the challenging problem of quantization without
gauge fixing (keeping the diffeomorphism and Hamiltonian constraints).

To end this section we want to mention the paper by Husain and
Winkler~\cite{Husain:2004yz} that considers the quantization of the
spherically symmetric gravity plus massless scalar fields after a
gauge fixing that reduces the theory to a model with a single
constraint generating radial diffeomorphisms. Though they do not work
in the framework of loop quantum gravity proper, they employ a type of
polymeric quantization somehow inspired in LQG. In this quantization
there are operators corresponding to the curvature that can be used
to discuss issues related to singularity resolution (at the dynamical
level). From a technical point of view an interesting detail in this
work is the use of Painlev\'e--Gullstrand coordinates that avoid the
necessity to consider the interior and and exterior problems
separately.

%================================================================
\newpage

\section{Conclusions, Open Issues and Future Problems}
\label{sect6}

As we have seen, the quantization of midisuperspace models has been --
and still is -- a very useful test bed to understand many different
methods and techniques proposed for the quantization of general
relativity. In this respect the models described in this Living Review
will be certainly of use in the near future, in particular to explore
new avenues to the quantization of gravity both in the
geometrodynamical approach and in loop quantum gravity. These models
have also been used with some success to extract qualitative physical
results that are expected to be present in full quantum gravity (black
hole mass quantization, Hawking radiation, unitarity of quantum
evolution, microcausality, singularity resolution, semiclassical
limit,\ldots). Certainly a complete full quantization of the symmetry
reductions described here will shed light on many methods and
approaches and may be useful to arrive at a fully functioning theory
of quantum gravity.

\bigskip

\noindent Among the open problems that we think should be considered we provide
the following list:

\begin{itemize}

\item From a purely mathematical point of view it would be interesting
  to study how midisuperspace models sit inside full superspace (both
  in metric and connection variables) in the spirit of analysis carried
  out by Jantzen for minisuperspaces~\cite{Jantzen:1979}.

\item Advance in the study of one-Killing reductions (or equivalently
  2+1 gravity coupled to matter fields). It is important to find
  out if these systems can be treated perturbatively after carefully
  introducing the necessary boundary terms in the action. From the
  non-perturbative point of view it is necessary to obtain the
  concrete form of the bounded Hamiltonians for open spatial
  topologies and understand their physical implications.

\item Give a complete and unified Hamiltonian treatment of all the
  midisuperspaces that admit a general two-dimensional spatial
  isometry group, generalizing in this way the two-Killing vector
  reductions of general relativity considered so far. Special
  attention should be paid to the analysis of non-compact spatial
  topologies and general polarization.

\item Obtain physical predictions from these models.

\item It is important to attempt the quantization of two-Killing
  vector reductions coupled to different types of matter fields beyond
  massless scalars. Though there are some papers dealing with scalar
  and electromagnetic fields there is still a lot of work to do. It
  must be said, however, that it may be actually be very difficult to
  get exact solutions to them.

\item Using the non-perturbative quantizations provided by Korotkin,
  Nicolai and Samtleben \cite{Nicolai:1996pd} to obtain physical
  predictions and check if
  expected quantum gravitational phenomena do actually occur. Among
  these we would suggest to study microcausality in these
  models. Another interesting issue may be the coupling of matter
  fields -- in particular massless scalars. If they can be described
  by using Fock spaces their particle-like quanta might be used as
  quantum test particle to explore quantum geometry and the emergence
  of a classical spacetime.

\item It would be interesting to provide consistent and complete
  quantizations inspired in loop quantum gravity for Einstein--Rosen
  waves and Gowdy models -- for which the available results are rather
  incomplete --. In particular it is important to go beyond the current hybrid
  formulations to purely polymeric ones.

\item Complete the quantization of gravitational plane waves.

\item Advance in the geometrodynamical quantization of spherically
  symmetric gravitational systems coupled to matter. In particular in
  those situations where the reduced phase spaces are infinite
  dimensional. It would be desirable to reach a level of mathematical
  rigor on par with the one customary within loop quantum
  gravity.

 \item Complete the program started by Gambini, Pullin and
 collaborators to understand spherically symmetric gravitational
 systems (with and without matter) in loop quantum gravity.

\item Answer the following questions in the midisuperspace setting:
  Does loop quantum gravity predict the mass quantization of black 
  holes? Is it possible to describe Hawking radiation in this framework?

\item Understand the fate of classical singularities both cosmological
  and in black holes. In the case of black holes there are claims of
  singularity resolution both in the geometrodynamical approach and in
  loop quantum gravity. One should understand the possible
  relationship between both approaches.

\end{itemize}

We expect that a lot of progress on these issues will happen in the
near future. We will report on them in future updates of this Living
Review.

%================================================================
\newpage

\section{Acknowledgements}
The authors want to thank Daniel G\'omez Vergel and the anonymous referee for many
interesting comments. This work has been supported by the Spanish
MICINN research grants FIS2008-03221, FIS2009-11893 and the
Consolider-Ingenio 2010 Program CPAN (CSD2007-00042).

%================================================================

\end{document}